# Long-range Phase Coherence and Tunable Second Order $\varphi_0$-Josephson Effect in a Dirac Semimetal 1T-PtTe2

Pranava K. Sivakumar[1*], Mostafa T. Ahari[2], Jae-Keun Kim[1], Yufeng Wu[1], Anvesh Dixit[1], George J. de Coster[3], Avanindra K. Pandeya[1], Matthew J. Gilbert[2,4] and Stuart S. P. Parkin[1*].

1. Max Planck Institute of Microstructure Physics, 06120 Halle (Saale), Germany
2. Materials Research Laboratory, The Grainger College of Engineering, University of Illinois, Urbana-Champaign, Illinois 61801, USA
3. DEVCOM Army Research Laboratory, 2800 Powder Mill Rd, Adelphi, Maryland 20783, USA
4. Department of Electrical Engineering, University of Illinois, Urbana-Champaign, Illinois 61801, USA

\* - Corresponding Authors

email: sivakumar@mpi-halle.mpg.de, stuart.parkin@mpi-halle.mpg.de

## Abstract

Superconducting diode effects have recently attracted much attention for their potential applications in superconducting logic circuits. Several mechanisms such as magneto-chiral effects, finite momentum Cooper pairing, asymmetric edge currents have been proposed to give rise to a supercurrent diode effect in different materials. In this work, we establish the presence of a large Josephson diode effect in a type-II Dirac semimetal 1T-PtTe2 facilitated by its helical spin-momentum locking and distinguish it from extrinsic geometric effects. The magnitude of the Josephson diode effect is shown to be directly correlated to the large second-harmonic component of the supercurrent. We denote such junctions, where the relative phase between the two harmonics can be tuned by a magnetic field, as 'tunable second order $\varphi_0$-junctions'. The direct correspondence and tunability of the second harmonic supercurrents and the diode effect in 1T-PtTe2 junctions at relatively low magnetic fields makes it an ideal platform to study the Josephson diode effect and higher order supercurrent transport in Josephson junctions.

# Introduction

Over the past few years, there has been much interest in creating and understanding superconducting and Josephson diodes, from both a fundamental and technological perspective[1-10]. These devices exhibit non-reciprocal superconducting critical currents and allow for unidirectional propagation of supercurrents and normal currents in the opposite direction, which is quite promising for the creation of novel low dissipative technologies. The observation of a supercurrent diode effect requires the breaking of both inversion and time-reversal symmetries (TRS)[11], which makes it also interesting in itself as a useful 'tool' providing insights into a material's properties in the superconducting state such as the nature of spin-orbit coupling[8] and in the determination of a chiral superconducting state that breaks time-reversal symmetry[12].

In this paper, we perform a detailed study of the Josephson diode effect (JDE or $\Delta I_c$) in a transition metal dichalcogenide and Dirac semimetal system (1T-PtTe$_2$) in different current and magnetic field geometries. This allows for distinguishing between intrinsic contributions to the JDE arising from the band structure and extrinsic junction geometric effects and establish the presence of helical spin-momentum locking in the system. The supercurrent behavior in the junction is studied in detail by considering a current-phase relationship (CPR) with a second harmonic term that we refer to as a 'tunable second order $\varphi_0$-junction' CPR. The observations from this CPR are verified by measuring the evolution of critical currents in PtTe$_2$ junctions in the presence of a magnetic flux and a magnetic field that is needed to induce the JDE. These measurements are used to provide direct evidence that the oscillations in $\Delta I_c$ are second harmonic in nature with nodes occurring at every half-magnetic flux quantum $\left(\frac{\Phi_0}{2}\right)$ and that the magnitude of $\Delta I_c$ is closely related to the magnitude of second harmonic supercurrents in the system and a phase difference ($\delta$) between the first and second harmonic components, as predicted from the CPR. This CPR combined with the tunability of $\delta$ with a magnetic field provides the interesting possibility of controlling the relative magnitudes and direction of first- and second-harmonic supercurrents. Finally, the role of the helical spin-momentum locked topological states in the formation of high transparency interfaces and phase coherent higher order Andreev reflections in PtTe$_2$ junctions that leads to the presence of a strong second harmonic term and hence a large JDE in the system is discussed.

# Results

## Lateral Josephson junctions of PtTe$_2$

1T-PtTe$_2$ is an air stable two-dimensional Van der Waals transition metal dichalcogenide (TMDC) that crystallizes in the centrosymmetric $P\bar{3}m1$ crystal structure [Fig. 1(a)]. Though 1T structures in which the transition metal atom has an octahedral coordination are centrosymmetric down to the monolayer limit, they have local inversion symmetry breaking within a single layer at the chalcogenide sites, giving the transition metal atom a $D_{3d}$ point group symmetry and the chalcogenide atom a $C_{3v}$ point group symmetry. This local inversion symmetry breaking gives rise to a series of band inversions and topological surface states[13] in these materials along with Rashba spin-orbit coupling of equal magnitude on each of the chalcogenide layers within each monolayer, with the top and bottom chalcogenide atomic layers having opposite spin-orbit coupling strengths. These differences in point group symmetry are predicted to give rise to a layer dependent 'local Rashba effect' with helical spin-momentum locking of opposite helicities on alternating chalcogenide layers as has been observed through spin-polarization measured through spin- and angle-resolved photoemission measurements in certain group-X transition metal dichalcogenides including PtTe$_2$[14], but also, for example, PtSe$_2$[15] and PdTe$_2$[13] as well as in the cuprate superconductor Bi$_2$Sr$_2$CaCu$_2$O$_{8+x}$[16]. This helical spin-momentum locking in 1T structures is analogous to Ising spin-momentum locking in 2H TMDCs[17,18]. In the case of PtTe$_2$, Dirac cone-like dispersions with helical spin texture and net spin polarization have been observed near the Fermi level through angle- and spin-resolved photoemission[14].

Thin PtTe$_2$ flakes are exfoliated from a single crystal on a Si/SiO$_2$ substrate and lateral Josephson junctions are fabricated by depositing Ti/Nb/Au electrodes on top of them with varying separations as described in the Methods section. The normal state transport characteristics of PtTe$_2$ are provided in Supplementary Note 2. The optical image of one such PtTe$_2$ flake (shown in Supplementary Note 1) of around 17.5 nm thickness (33-34 layers) with multiple lateral Josephson junctions of varying separations (L1-L4) is shown in the inset to Fig. 1(b) which also shows the defined Cartesian coordinate axes for the device. The direction of current bias in these devices is fixed along the x-axis. The shortest separation between the niobium electrodes is around 390 nm, and the device is roughly 5 μm in width. The results presented in the main text are from this device (referred to as L1 hereafter) unless specified otherwise. The resistance of this junction is measured with a small current as the junction is cooled down [Fig. 1(c)]. A drop in resistance is observed around 4.5 K corresponding to the superconducting transition of the Nb electrodes and another drop at 2.7 K [Fig. 1(c) inset], below which the junction becomes fully superconducting $(T_J)$.

After cooling the sample to the base temperature of the dilution refrigerator(20 $mK$), current-voltage curves are measured in zero field. The critical currents on sweeping the current from zero bias in the positive ($I_c^+$) and negative ($I_c^-$) directions are obtained at zero magnetic field [Fig. 1(d)] and a negligible difference in their magnitude ($\Delta I_c = I_c^+ - |I_c^-|$) or JDE is observed. As the in-plane magnetic field perpendicular to the direction of current $(B_y)$ is increased, the appearance of a non-zero $\Delta I_c$ is seen as shown in Fig. 1(e). Previously, such a $\Delta I_c$ has also been observed in similar lateral junctions formed with 1T-NiTe$_2$[4], a transition metal dichalcogenide material with the same crystal symmetry and large spin-orbit splitting close to the Fermi level. The JDE has been attributed to the finite momentum Cooper pairing[11] induced by either the topological surface states with large Rashba spin-splitting or the Meissner screening currents within the electrodes.

It is important to note that in contrast to NiTe$_2$ junctions used in previous studies[4], the width ($W$, lateral dimension perpendicular to the direction of current flow) of the PtTe$_2$ flake forming the Josephson junction L1 is comparable to the Josephson penetration depth $(W \sim \lambda_J)$. In this limit, the effect of current-induced magnetic field, also known as 'self-field effect' (SFE) becomes significant and the geometry of the current source configuration can play a significant role in the current distribution across the junction. The SFE modifies the critical current of the junction, which can result in skewed Fraunhofer pattern under an out of plane magnetic field. A self-consistent treatment of the wide Josephson junction as described in Supplementary Note 3 is used to simulate the properties of the junction. It is shown in Supplementary Note 4 that in such junctions, it is possible to obtain extraneous JDE just by choosing the current bias electrodes to be on the same side and that this can be avoided by choosing a 'criss-crossed' current bias geometry that gives a rather uniform current distribution across the junction and minimizes the effect of the self-field. Detailed discussion on SFE and the extrinsic JDE resulting from it is provided in Supplementary Note 4. The remarkable match between our experimental data and simulations in both bias configurations reinforces the validity of our supposition. All measurements henceforth, presented on lateral junctions of PtTe$_2$ to determine the spin-momentum locking, were carried out in the criss-crossed geometry to minimize the influence of SFE.

## Helical spin-momentum locking induced JDE in PtTe$_2$

As stated earlier, PtTe$_2$ has helical spin-momentum locked states close to the Fermi level[14]. This helical spin-momentum locking is expected to give rise to a finite-momentum Cooper pairing (FMCP) and a $\Delta I_c$ in the presence of an in-plane magnetic field perpendicular to the current $(B_y)$[4,11]. In the junction geometry used, finite momentum Cooper pairing can also arise due to the Meissner screening currents in the superconducting electrodes[4,7,19,20] but we argue in the discussion section that this effect is negligible and contributes very little to the observed diode effect based on the transparency of our junctions. To establish the presence of helical spin-momentum locking and rule out the presence of three-dimensional spin-orbit coupling in PtTe$_2$, two different configurations of devices were used: lateral junction (L1) as discussed above and a vertical Josephson junction (VJJ) with a PtTe$_2$ flake sandwiched by NbSe$_2$ flakes on top and bottom (that is labelled V1). Fig. 2(a) shows $I_c^+$ and $I_c^-$ as a function of $B_y$ in L1 and the corresponding $\Delta I_c$ is shown in Fig. 2(b). It can be seen that $\Delta I_c$ increases linearly with $B_y$ at low fields and then starts to fluctuate and decrease non-monotonously. This is due to an additional magnetic flux to the sample from $B_y$ that can either be due to the finite sample thickness or a tiny misalignment or flux focusing effect that leads to an additional phase difference across the electrodes. For this reason, we use the Fraunhofer interference pattern to locate the exact critical current maxima where the net out-of-plane magnetic flux is zero and extract the value of the diode effect due to $B_y$ only (Refer Supplementary Note 7). This method of extracting the diode effect helps avoid any pitfalls due to magnetic flux through the sample. The effect of flux focusing and the determination of the flux focusing factor ($\Gamma$) is described in Supplementary Note 8.

On measuring the critical currents $I_c^+$ and $I_c^-$ of L1 at 20 mK as a function of the in-plane magnetic field angle [Fig. 2(d)], it is seen that $\Delta I_c$ is maximized when the magnetic field is applied perpendicular to the direction of current $(B_y)$ and vanishes when the magnetic field is along the direction of current $(B_x)$. $\Delta I_c$ also decreases as a function of temperature at higher temperatures with a quadratic $(T - T_J)^2$ dependence as expected for a finite momentum Cooper pairing scenario[4,11], as shown in Fig. 2(c). Furthermore, there is no clear evidence of a $\Delta I_c$ in vertical Josephson junctions of PtTe$_2$ with an in-plane magnetic field along different directions (Refer Supplementary Note 5), as opposed to that in vertical junctions of T$_d$-WTe$_2$ where a clear $\Delta I_c$ is observed[21]. Thus, this result shows the absence of net spin-momentum locking or any other finite momentum pairing mechanism when the current flows along the c-axis of PtTe$_2$. All the above results together point to the existence of a two-dimensional helical spin-momentum locking in PtTe$_2$. The absence of any significant contribution to the JDE due to geometric inversion asymmetry of the PtTe$_2$ flake is shown in Supplementary Note 6.

## Tunable second-order supercurrents and Current-Phase relationship (CPR) induced by Finite Momentum Cooper Pairing (FMCP) in PtTe$_2$

Having established the existence of a helical spin-momentum locking in PtTe$_2$, the evolution of the Fraunhofer pattern in lateral PtTe$_2$ junctions in the presence of $\Delta I_c$ is studied to gain insight into the CPR of the system. While superconducting quantum interference devices (SQUIDs) are the preferred platforms to deduce the current-phase relationship in a system, Josephson junctions have the advantage that the distribution of supercurrents in the system may also be obtained by analyzing the Fourier transform of the Fraunhofer pattern. The Fraunhofer patterns for the critical currents, $I_c^+$ and $I_c^-$ are measured as the function of the magnetic flux $\Phi$ along the z-direction, under various $B_y$ is shown in Fig. 3(a)-(d) after correcting for flux focusing effects[22,23] and the finite thickness effect[24,25] of the sample as discussed in detail in Supplementary Notes 7 and 8. When $B_y = 0$, $I_c^+$ and $I_c^-$ lie on top of each other leading to a negligible $\Delta I_c$ and the period of oscillations is close to a single magnetic flux quantum $\left(\Phi_0 = \frac{h}{2e}\right)$ as expected [Fig. 2(f)]. As $B_y$ is increased in the negative direction to $-8$ mT and the Fraunhofer pattern is measured again [Fig. 3(a)], it is observed that the central maxima of $I_c^-$ increases slightly in magnitude while the magnitude of the central peak of $I_c^+$ starts to decrease. As the magnetic field is increased further from $-12$ mT to $-24$ mT, [Fig. 3(b-d)] the central peak of $I_c^-$ doesn't decrease much in magnitude while the magnitude of the central peak of $I_c^+$ has a sharp decrease in the middle leading to the formation of a sharp noticeable dip in critical current where maximum $\Delta I_c$ is observed. It is to be noted that in these experiments, the roles of $I_c^+$ and $I_c^-$ are reversed when $B_y$ is swept in the opposite direction and that corresponds to $I_c^+(B_y, B_z) = -I_c^-(-B_y, -B_z)$, indicating that the total time-reversal symmetry of the system is maintained and there is no other external sources of magnetic flux, like vortices trapped in the system (Refer Supplementary Note 9).

The appearance of $\Delta I_c$ in PtTe$_2$ can be understood using a simple model starting from a general current-phase relationship (CPR) written as a Fourier series of sine functions, which includes higher harmonics and additional phase shifts $\varphi_n$ that may be present in the system when time-reversal symmetry is broken.

$$I(\varphi) = \sum_{n=1}^{\infty} I_n \sin(n\varphi + \varphi_n) \qquad (1)$$

This CPR can be expanded up to the second order as higher order supercurrents contribute negligibly to the total current. This gives us:

$$I(\varphi) = I_1 \sin(\varphi + \varphi_1) + I_2 \sin(2\varphi + \varphi_2) \qquad (2)$$

Certain well-known cases of unconventional CPR can be derived from this generic CPR. For example, having $\varphi_1 = \varphi_2 = 0$ in Eq.2 gives a CPR that contains only the first and second harmonic terms without any additional phases corresponding to typical $\varphi$-junctions[26,27] with a skewed current-phase relationship. In the case where $I_2 = 0$ in Eq. (2), it gives rise to anomalous Josephson junctions or $\varphi_0$-junctions with a sinusoidal current phase relationship shifted from zero by a phase $\varphi_1$. Such CPRs have been observed typically in ferromagnetic Josephson junctions and systems with high spin-orbit coupling[28-33]. Now, without any loss of generality $\varphi$ may be replaced with $(\varphi - \varphi_1)$ and the CPR can be rewritten as:

$$I(\varphi) = I_1 \sin\varphi + I_2 \sin(2\varphi + \delta) \qquad (3)$$

introducing $\delta = \varphi_2 - 2\varphi_1$, the relative phase between the first and second harmonic terms. We shall call such a Josephson junction with a CPR as that in equation (3) as a 'tunable second order $\varphi_0$-junction'. This CPR is identical to that derived from the Ginzburg-Landau formalism[4], in which $\delta$ corresponds to the phase shift induced by a finite momentum Cooper pairing in the system. The phase, $\delta$, may be controlled by an in-plane Zeeman field perpendicular to the direction of current ($\delta \propto B_y$). We note that, in addition to NiTe$_2$[4], similar CPRs have been used to explain the presence of a $\Delta I_c$ in InAs-based superconducting junctions [9,34] and InSb nanowire junctions[35].

$\Delta I_c$ may be determined by examining $I_c^+$ and $I_c^-$ from the CPR in Eq. (3). For non-sinusoidal CPRs, such as that in Eq. (3), that are composed of higher order Fourier harmonics, critical currents may not occur at $\varphi = \pm\frac{\pi}{2}$ and need to be solved numerically to obtain the exact values of $I_c^+$ and $I_c^-$ for different values of $\delta$. We obtain the critical currents as

$$I_c^+(\Phi, \delta) = \max_{\varphi}[I_{tot}(\varphi, \Phi, \delta)]$$

$$I_c^-(\Phi, \delta) = \min_{\varphi}[I_{tot}(\varphi, \Phi, \delta)]$$

where $I_{tot}(\varphi, \Phi, \delta)$ denotes the total current given by

$$I_{tot}(\varphi, \Phi, \delta) = \frac{1}{W}\int_{-\frac{W}{2}}^{\frac{W}{2}} dy\ I\left(\varphi + 2\pi\frac{\Phi}{\Phi_0}\frac{y}{W}\right)$$

$$= \left(I_1 \sin\varphi + I_2 \cos\left(\pi\frac{\Phi}{\Phi_0}\right)\sin(2\varphi + \delta)\right)\frac{\sin(\pi\Phi/\Phi_0)}{\pi\Phi/\Phi_0}$$

For $\frac{I_2}{I_1} \ll 1$, the critical currents $I_c^{\pm}$ occur at $\varphi = \pm\frac{\pi}{2}$, so that

$$\Delta I_c(\Phi, \delta) = -2I_2(B)\sin\delta\ \frac{\sin\left(\frac{2\pi\Phi}{\Phi_0}\right)}{\frac{2\pi\Phi}{\Phi_0}} \tag{4}$$

where $I_n(B) = I_n(0)\left(1 - \frac{B^2}{B_c^2}\right)^n$ accounts for the suppression in the critical current components due to $B$. We note that $\Delta I_c$ can serve as a probe for second harmonic supercurrent in the junction because the first harmonic term does not have any direct contribution to $\Delta I_c$. From Eq. (4), we infer three important conclusions on the nature of the CPR and $\Delta I_c$. First, we find that the existence of a second-harmonic term ($I_2 \neq 0$) and $\delta \neq n\pi, n \in \mathbb{Z}$ is necessary for the existence of a non-zero $\Delta I_c$. So, the presence of a $\Delta I_c$ acts as an indicator for the existence of a second harmonic term in the current-phase relationship while the converse is not true. Second, control over $\delta$ leads to the possibility of tuning the relative phase between the first and second harmonic components, which leads to control of specific harmonics. For example, substituting $\delta = \pi$ in Eq. (3) gives:

$$I(\varphi)_{\delta=\pi} = I_1 \sin\varphi + I_2 \sin(2\varphi + \pi) = I_1 \sin\varphi - I_2 \sin 2\varphi \tag{5}$$

In this CPR, the first and second harmonics of supercurrents have opposite signs and can flow in opposite directions. Hence by tuning the magnetic flux and choosing a suitable value of $\delta$ and $\varphi$, the magnitude and flow direction of pure second or first order supercurrents across the junction can be controlled. Third, the magnitude of $\Delta I_c$ is modulated by $\sin \delta$, which implies that $\Delta I_c$ reaches its largest magnitude when $\delta = \pm \frac{\pi}{2}$. In a system with FMCP such as 1T-PtTe$_2$, $\delta$ may be tuned precisely with an in-plane magnetic field $(B_y)$. Assuming that $I_2$ and $I_1$ are both positive, the value of $B_y$ at which $\Delta I_c$ reaches the maximum (minimum) value $\Delta I_c^{\max}$ ($\Delta I_c^{\min}$) corresponds to $\delta = -\frac{\pi}{2}$ ($\delta = \frac{\pi}{2}$) from Eq. (4). Using the value of $\Delta I_c^{\min}$ in Eq. (4), we observe that the magnitude of second harmonic supercurrent flowing through the junction is $I_2(B_y) = -\frac{\Delta I_c^{\min}}{2}$, in the limit of $\Phi$ going to zero. For junction L1, the minimum value of $\Delta I_c$ is around $-34$ µA at $B_y = -24$ mT [Fig. 3(l)], this would produce $I_2(-24 \text{ mT}) \approx 17$ µA and the actual value of $\frac{I_2(0)}{I_1(0)} \approx 0.37$. The value of $\frac{I_2}{I_1}$ obtained from this analysis is larger than that measured in some semiconductor junctions with high transparency such as Sn-InSb nanowire junctions[35] and comparable to that observed in Al-InAs planar Josephson junctions[36]. Calculating the Josephson diode efficiency $\eta = \frac{\Delta I_c}{I_c^+ + |I_c^-|}$ for this junction at maximum $\Delta I_c$ gives a value of around 32 % at $-24$ mT, which is one of the largest values reported so far (Refer Supplementary Note 15 for comparison).

In order to corroborate the validity of values estimated from the model, the Fraunhofer patterns for $I_c^+$, $I_c^-$ and, consequently, $\Delta I_c$ corresponding to different values of $\delta$ are simulated with $\frac{I_2}{I_1} \approx 0.4$ in Figs. 3(e)-(h) and Figs. 3(m)-(p), respectively. We observe that the CPR captures the main features of the experimental data such as the magnitude of $\Delta I_c$ and the oscillation period of it. Some additional features for small $B_z$ such as lifted nodes in $\Delta I_c$ and the formation of a dip in the critical currents at zero $B_z$ can be captured by introducing an additional term to the phase difference such that $\varphi \to \varphi + \beta \frac{I_{\text{tot}}}{I_1} \left|\frac{y}{W}\right|$, which is due to a small remnant self-field in the junction. In Supplementary Note 3, we provide a derivation of $\beta$ and show that $\beta = \frac{1}{2} \left(\frac{W}{\lambda_J}\right)^2$. This term directly shows the influence of junction geometry (wide junction) on the phase gradient. The corresponding CPRs for negative and positive $B_y$ are shown in Fig. 3(q) and Fig. 3(r) respectively, where the non-reciprocal nature of the critical currents can be seen clearly. The details of the simulations are relegated to Supplementary Note 3. It is seen that the features of $I_c^+$ and $I_c^-$ from the simulation are in qualitative agreement with the experimentally measured curves. We note that the features of the simulation that are also observed in experiment such as the sharp peak in $I_c^+$ around $-12$ mT and the observed dip in $I_c^+$ beyond $-16$ mT are quite sensitive to the value of $\frac{I_2}{I_1}$ and the origin of these features are reflected in the calculated CPR curves [Fig. 3(q)]. It can be seen in the CPR curves that as $|B_y|$ is increased, the critical current in the negative direction ($I_c^-$) first increases in magnitude and then starts to decrease gradually with a steady shift in the value of $\varphi$ at which it occurs, while in the case of $I_c^+$ there is initially a gradual decrease in its value as $B_y$ is increased with a shift in the value of $\varphi$.

Further, the evolution of $I_c^+$ and $I_c^-$ as a function of $B_y$ plotted in Fig. 4(a) is well replicated by the corresponding simulation presented in Fig. 4(b). The absence of nodes in the experimental observation of $\Delta I_c$ versus $\delta$ in Fig. 4(a) and Fig. 4(c) can be explained by the presence of a small magnetic flux induced by $B_y$ as shown in Supplementary Note 13. At this point, a digression on the effect of the in-plane magnetic flux due to $B_y$ is warranted. The effective in-plane cross sectional area of the junction including the London penetration depth of the two superconducting electrodes is around $1.1725 \times 10^{-14} \text{m}^2$, which gives the effective magnetic field needed to induce a single magnetic flux quantum in the junction is around 176 mT and the in-plane fields that we use in our study is much lower than this ($0 - 50$ mT) to create any magnetic flux-induced oscillations. Moreover, the evolution of $I_c$ with $\delta$ is quite different from what is expected for a typical flux-induced Fraunhofer pattern. There is an inherent asymmetry between $I_c^+$ and $I_c^-$ that arises due to $\delta$ as shown by the simulations in Fig. 4(b) and observed experimentally in Fig. 4(a). Rather than having nodes, the evolution of $I_c$ with $\delta$ has oscillations that decay slowly with no nodes in the critical current. Hence, we conclude that flux-induced Fraunhofer interference effects due to $B_y$ in our junctions are not relevant to the observed diode effect.

The intimate correlation between the experimentally observed features [Fig. 4(a)] and the simulation [Fig. 4(b)] demonstrates the accuracy of the assumed CPR. Fig. 4(c) shows $\Delta I_c$ as a function of $B_y$ as derived from the Fraunhofer interference pattern at zero net magnetic flux. $\Delta I_c$

deviates from the expected sinusoidal behavior and increases in magnitude linearly with $B_y$ till $\pm 24$ mT and then decreases linearly towards zero. This behavior can also be reproduced successfully in the simulations by tuning the $\left|\frac{I_2}{I_1}\right|$ ratio as shown in Fig. 4(d). While $\Delta I_c$ vs $B_y$ remains sinusoidal for lower values of $\left|\frac{I_2}{I_1}\right|$, it gradually turns triangular for larger values of $\left|\frac{I_2}{I_1}\right|$ for wide junctions. This deviation of $\Delta I_c$ from the sinusoidal behavior expected from Eq. (4) also confirms the presence of large second-harmonic supercurrents.

One of the main observations from Eq. (4) is that $\Delta I_c$ is expected to oscillate with the magnetic flux $\Phi$ with nodes at every half-flux quantum $\left(\frac{\Phi_0}{2}\right)$ due to the presence of the second-harmonic term in the CPR. The oscillations in $\Delta I_c$ as a function of $\Phi$ at $B_y = 20$ mT are presented in Fig. 4(e). Though the oscillations are expected to vanish, the first few nodes in $\Delta I_c$ are lifted from their zero position, which is similar to that observed in the Fraunhofer patterns for $I_c^+$ and $I_c^-$ [Fig. 2(f)]. Lifted nodes in Fraunhofer patterns may be due to several mechanisms, such as the junction geometry[37], current asymmetry, or a remnant of an unconventional CPR due to topological superconductivity[38,39]. The lifted nodes encountered in our case can be accounted for in the simulations by the presence of a self-field related to the junction geometry that results in non-zero $\beta$. $I_c^+$ simulated for different values of $\beta$ is shown in Fig. 4(g). Interestingly, it can be seen that the appearance of a dip in $I_c^+$ upon increasing $\delta$ [Fig. 4(f)] can be captured by increasing $\beta$, suggesting the intimate correlation between these two parameters as assumed. The variations in the magnetic flux at which these features can be observed experimentally is due to a varying flux-focusing factor close to zero magnetic flux[23] and the first lifted node in $I_c^+$, which masks the dip close to the first magnetic flux quantum. The simulation of lifted nodes in $\Delta I_c$ due to a non-zero $\beta$ are presented in Supplementary Note 13. The oscillations are observed to have nodes that are roughly spaced every $\left(\frac{\Phi_0}{2}\right)$ strongly indicating that the major component in $\Delta I_c$ is close to second harmonic as expected in a tunable second order $\varphi_0$-junction. The second harmonic component extracted from other junctions (L3 and L4) together with L1 (shown in Supplementary Notes 10-12) is observed to scale quadratically with the $I_c$ of the junctions, as shown in Fig. 5(f), which is expected from a Ginzburg-Landau analysis[4], further confirming our hypothesis. The junction L2 that is accidentally shorted by another flake and forms an asymmetric SQUID displays skewed, non-sinusoidal oscillations of the critical current, also indicating the presence of a large second-harmonic term in the CPR of the junction[40] (Refer Supplementary Note 14).

## Discussion

Now, we turn our attention to the physical origin of a large second-harmonic term in $PtTe_2$ Josephson junctions. The presence of such a large second harmonic component in a Josephson junction with a large electron density made over such long separations is quite unanticipated as it indicates a large transparency of the interface between the superconducting electrodes and $PtTe_2$, and the presence of resonant Andreev bound states due to phase coherent transport across the junction[41]. The band structure studies of multilayer $PtTe_2$ have shown the formation of high mobility Fermi pockets at the Fermi level[42,43]. This means that the difference between the Fermi level and the edges of these bands is small. As a result, a relatively large external magnetic field can lead to significant warping and even the (dis)appearance of the Fermi pockets, i.e., a Lifshitz transition[44], which could strongly affect the density of states and velocity at the Fermi level. A large density of states available for Cooper pair transfer, as a result, can enhance the critical currents associated with single ($I_1$) and double Cooper pair transfer ($I_2$) in the junction.

Moreover, we note that large transparencies and consequently higher harmonics are observed in high-mobility semiconducting[45] and semi-metallic junctions[46] with low electron densities and pristine interfaces but it is not so in metallic junctions due to short mean-free paths and scattering at the interface. The mean-free path ($l_e$) reported in literature for single crystals of $PtTe_2$ is around 180 nm in the $ab$ plane[47], which is large in comparison with a normal metal with similar carrier densities, where it is typically of the order of $1 - 50$ nm[48]. Josephson junctions made with metallic barriers have large critical currents owing to large density of states available for Cooper pair tunneling. As discussed earlier, since the second harmonic supercurrent ($I_2$) scales roughly as the square of $I_c$, $I_2$ is typically larger in metallic Josephson junctions with larger $I_c$ as in our case (~73 µA) as compared to semiconductor Josephson junctions with much lower electron densities[10] (~1 µA). This enhancement in $I_c$ is advantageous for easier and clear observation of higher order effects in the junction like the oscillations in $\Delta I_c$ that we observe.

The superconducting coherence length $\left(\xi = \frac{\hbar v_F}{\pi \Delta_0}\right)$ of L1 is calculated to be around 200 nm at zero temperature using the average value of $v_F \approx 3.3 \times 10^5$ ms$^{-1}$ reported in literature[47] for PtTe$_2$. It is neither clearly in the short or long junction limit when compared to the junction separation (390 nm) and thus is not straight forward to determine the transparency of the junction by fitting $I_c(T)$ using a standard model for a long junction. Instead, the transparency of this junction is obtained by examining excess currents ($I_e$) that are obtained by linear extrapolation of the $I - V$ curve above the critical current back to zero voltage, as shown in Fig. 5(a). The existence of $I_e$ in a highly transparent junction with long-range phase-coherent Andreev reflections is explained by the Octavio-Tinkham-Blonder-Klapwijk (OTBK) model[49,50]. The $I_e$ for L1 at 20 $mK$ is around 9 µA which corresponds to a transparency of around 0.45.

The transparency of the other junctions L3 and L4, which are in the long junction limit ($d \gg \xi$), is obtained by fitting the critical current over the entire temperature range in the long junction limit[51,52] given by $I_c(T) = \eta \frac{aE_T}{eR_n}\left[1 - be^{\frac{-aE_T}{3.2 k_B T}}\right]$. $a$ and $b$ are fitting parameters. $E_T$ is the Thouless energy and $R_n$ is the normal state resistance. The details of the fit can be found in the SI. The extracted transparency from the fits for L3 and L4 (Fig. 5b and 5c) is around 0.436 and 0.428 respectively, which is consistent with the values from excess currents. The transparency of the junctions is not as large as that found in semiconductor junctions, which is closer to unity in ballistic transport. In the case of PtTe$_2$, the decreased transparency is due to the contribution of diffusive channels in parallel to transparent ballistic channels. The presence of relatively high transparency despite the ex-situ fabrication of junction interface can be attributed to the significant contribution of the states with helical spin-momentum locking to the transport of supercurrents. We conjecture that the states in PtTe$_2$ with a spin-momentum locked Dirac-like dispersion[14] can suppress normal reflections due to the reduced availability of spin conserving states. This leads to coherent Andreev processes over long distances leading to strong second harmonic supercurrents as has been reported previously in other topological systems[53-55] and depicted in Fig. 5g. For instance, an electron in PtTe$_2$ moving to the right with an up spin, incident on the surface of the superconductor can be reflected as a hole moving to the left into the same band with opposite spin, which corresponds to the Andreev reflection process. Whereas for the normal reflection process which conserves spin upon reflection, the availability of spin states for the reflected electron is strongly dependent on the incident angle, in a system with spin-momentum locked Dirac-like dispersion as shown previously[53] and can be highly suppressed for incidences close to the normal. Similar arguments have also been presented in another recent work on highly transparent Dirac semimetal MoTe$_2$ junctions[56].

The transparency of the junctions also provide important clues into the dominant mechanism of JDE in our junctions and determining the effect of trivial mechanisms such as Meissner screening currents[4,7,19,20] that can also lead to a JDE. Application of a magnetic field perpendicular to current can induce finite-momentum Cooper pairing in materials with helical spin-momentum locking. Finite momentum Cooper pairs can also be generated through Meissner screening currents in the superconducting electrodes that can also induce a JDE[7,19,20]. While it is hard to completely disentangle the JDE arising due to these two mechanisms, we can make arguments based on the junction transparency to establish the dominant role of helical spin-momentum locking in the observation of a large JDE. It is important to note that the JDE generated by Meissner screening currents is extremely sensitive to the transparency of the junction[7,57] and drops drastically with decrease in transparency. For junctions with transparency of 0.45, the maximum diode efficiency due to Meissner screening currents is predicted to be around 4%[7,57]. In our case, the experimentally observed diode efficiency is close to 32% for similar transparencies, which is much larger and cannot be accounted for completely by finite momentum Cooper pairs created by orbital effects only.

The large spin-orbit coupling effect at small magnetic fields as evidenced from the non-zero momentum of the Cooper pairs indicates a large $\left(\frac{g}{v_F}\right)$ ratio in PtTe$_2$, which in turn points to the presence of an extremely large $g$-factor (in the order of $10 - 100$) for the electrons in PtTe$_2$ that depends on the exact value of $v_F$ for the bands that contribute to the supercurrent transport. However, it is hard to precisely estimate the $g$-factor directly from the JDE. A discussion on the $g$-factor estimation from the JDE and its limitations are presented in Supplementary Note 16. It is to be noted that similarly large $g$-factors have been previously reported in other topological semimetals[58,59] and semiconductor heterostructures[60,61]. The large spin-orbit coupling and Zeeman splitting with small magnetic fields coupled with the strongly coherent higher order Cooper pair transport in PtTe$_2$, as evidenced by its large JDE, provides an interesting alternative platform to engineer topological superconductivity in planar Josephson junctions[62,63] as has been demonstrated before in Josephson junctions of HgTe[64] and InAs[65] quantum well structures. One of the major challenges in the current existing platforms for realizing topological superconductivity is the engineering of high quality interfaces[66,67]. The complete air stability of PtTe$_2$ and the states with strong helical spin-momentum locking allow for creation of high quality interfaces with superconductors without many complications.

## Conclusion

In summary, we have shown through measurements of $\Delta I_c$ that the Dirac semimetal 1T-PtTe$_2$ has a large JDE that arises from its helical spin-momentum locked states under a Zeeman field. While extrinsic effects such as SFE can be present in wide Josephson junctions that can also lead to a JDE, it is shown that such extrinsic effects can be suppressed by using a criss-crossed measurement geometry. The junctions are shown to behave as 'tunable second-order $\varphi_0$-junctions', in which the supercurrent transport can be tuned between Cooper pairs and Cooper quartets of charges $2e$ and $4e$ respectively, through the analysis of $\Delta I_c$ in the Fraunhofer interference pattern and comparison with the proposed $\delta$-dependent CPR. The simulated $\Delta I_c$ with a strong second harmonic term $\left(\frac{I_2}{I_1} \approx 0.4\right)$, as inferred from the CPR analysis well replicates the experimental behavior. $\Delta I_c$ is also shown to have nodes at every $\left(\frac{\Phi_0}{2}\right)$, further confirming the validity of the proposed $\delta$-dependent CPR. Besides being important for the observation of a JDE, this CPR has unique properties such as the controlling the relative phase difference between the two harmonics in the junction and controlling the relative direction of supercurrent flow by tuning the $(\delta, \varphi)$ phase space. Josephson junctions of topological materials have been explored largely in the context of topological superconductivity[64,65] and though the protection against backscattering offered by the topological states leading to higher order Andreev processes has been reported in junctions prior to this work[40,53,56], their role in the creation of a large JDE and controlling its magnitude has been unambiguously identified and explored in this work, making them more relevant in creating supercurrent diodes of much larger efficiencies. Moreover, they would also be useful in the study of the $4e$ Cooper quartet transport without the need for multiple superconducting terminals[68,69], which have also been predicted to be useful in the creation of parity protected superconducting qubits[70,71]. We would also like to note the recent observation of 4e supercurrents in superconducting quantum interference devices (SQUIDs) consisting of InAs-Al heterostructure[72].

# Methods

**Exfoliation**

Thin flakes of $PtTe_2$ were exfoliated from a single crystal of 1T-$PtTe_2$ (purchased from HQ Graphene) under ambient conditions on a Si/$SiO_2$ substrate using a Nitto adhesive tape (SPV 224) and standard exfoliation techniques. Very thin flakes of few layer thicknesses are hard to obtain due to stronger interlayer attraction present in $PtTe_2$. A thin flake of around 17.5 nm thickness with a relatively large area was identified with the help of an optical microscope and its thickness was determined using an atomic force microscope (AFM). This flake was then used to fabricate the Josephson junctions presented in the main text.

**Device fabrication**

The Josephson junctions were fabricated on this flake using electron-beam lithography. The substrate containing the flake was spin-coated at 4000 rpm with a positive resist AR-P 669.04 and annealed at $150\,^oC$ for 60 seconds followed by the same procedure for AR-P 679.03 (purchased from Allresist GmBH). The substrate was then exposed to the electron beam at 10 kV energy and developed using AR 600-56 for 90 seconds. After development and gentle ion milling to remove residual resist on top surface, superconducting electrodes Ti (2 nm) / Nb (40 nm) / Au (4 nm) substrate was sputtered on the substrate. The lift-off was performed by immersing the substrate in acetone overnight.

**Electrical Measurements**

Electrical measurements were performed in a Bluefors LD-400 dilution refrigerator with a bottom-loading probe and a base temperature of 20 mK. The fridge is equipped with RF and RC filters (from QDevil Aps) that help decrease the electron temperature during measurements. DC measurements were performed to obtain the current-voltage characteristics of the Josephson junctions. The current bias was applied through a Keithley 6221 current source and the voltage was measured using a Keithley 2182A nanovoltmeter. A two-dimensional superconducting vector magnet attached to the system was used to control the magnetic field and measure the oscillations in the interference under different in-plane magnetic fields.

# Figures

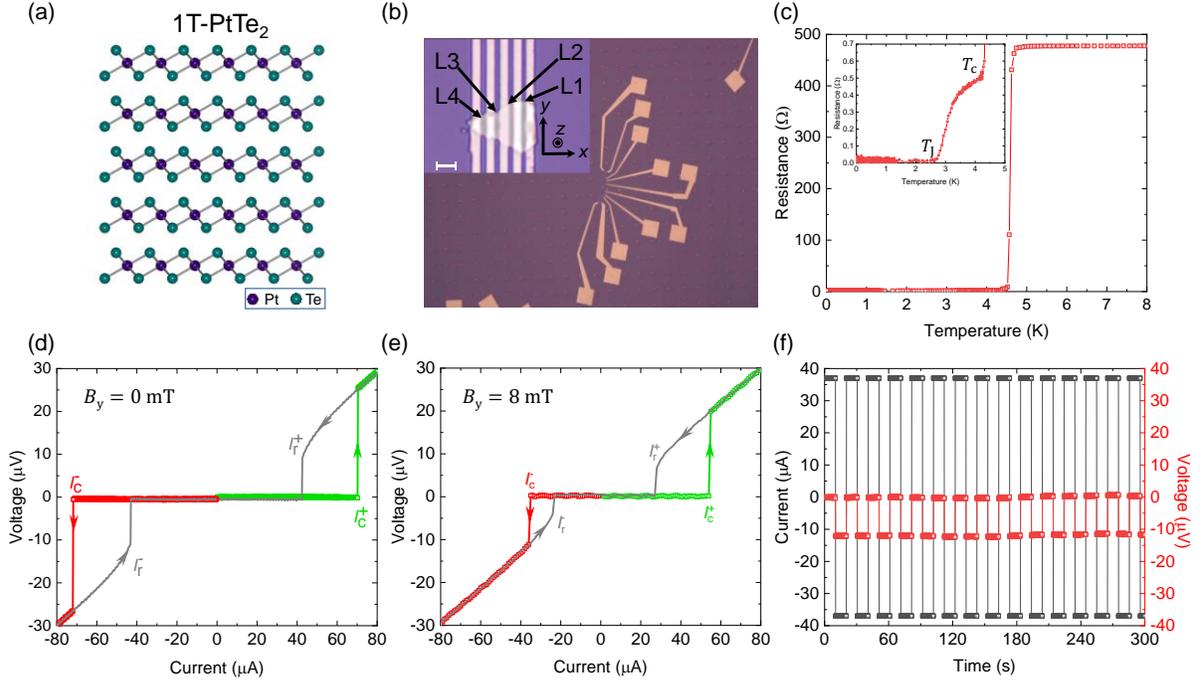

**Fig.1 Josephson junctions of PtTe₂ and non-reciprocal critical currents.** (a) A schematic of the 1T-PtTe₂ crystal, which shows the structure of five layers of platinum ditelluride with a trigonal prismatic coordination. Purple color indicates platinum atoms and green color indicates tellurium atoms. The structure comprises two dimensional, centrosymmetric layers of PtTe₂ stacked on top of each other separated by a Van der Waals gap. (b) An optical image of Josephson junction devices fabricated on a single 17 nm thick PtTe₂ flake. Inset shows a close up of devices with niobium electrodes with increasing separations labelled L1 to L4. The cartesian coordinate axes that are used are shown. The white scale bar represents 2 μm. (c) Resistance curve of junction L1 measured while cooling down in zero field. Two transitions at around 4.5 K and 2.7 K (inset) corresponding to the superconductivity of niobium ($T_c$) and the junction ($T_J$) are observed. (d) Current-Voltage characteristics of L1 measured in the absence of any external magnetic field after cooling down in zero magnetic field. The critical currents in the positive ($I_c^+$) and negative ($I_c^-$) directions are the same within the limit of error, making $\Delta I_c = 0$. The retrapping currents in both directions ($I_r^+$ and $I_r^-$) are also equal. (e) Current-Voltage characteristics of L1 measured in the presence of an 8 mT magnetic field applied along $B_y$. In addition to a suppression of the energy gap of the junction, we also observe that there is a significant difference in $I_c^+$ and $I_c^-$ leading to a $\Delta I_c$. (f) The non-reciprocal behavior of supercurrents measured under the same 8 mT magnetic field with a 37 µA current shows that the device is superconducting along one direction but resistive in the other direction. The switching was measured over a period of one hour and showed robust behavior.

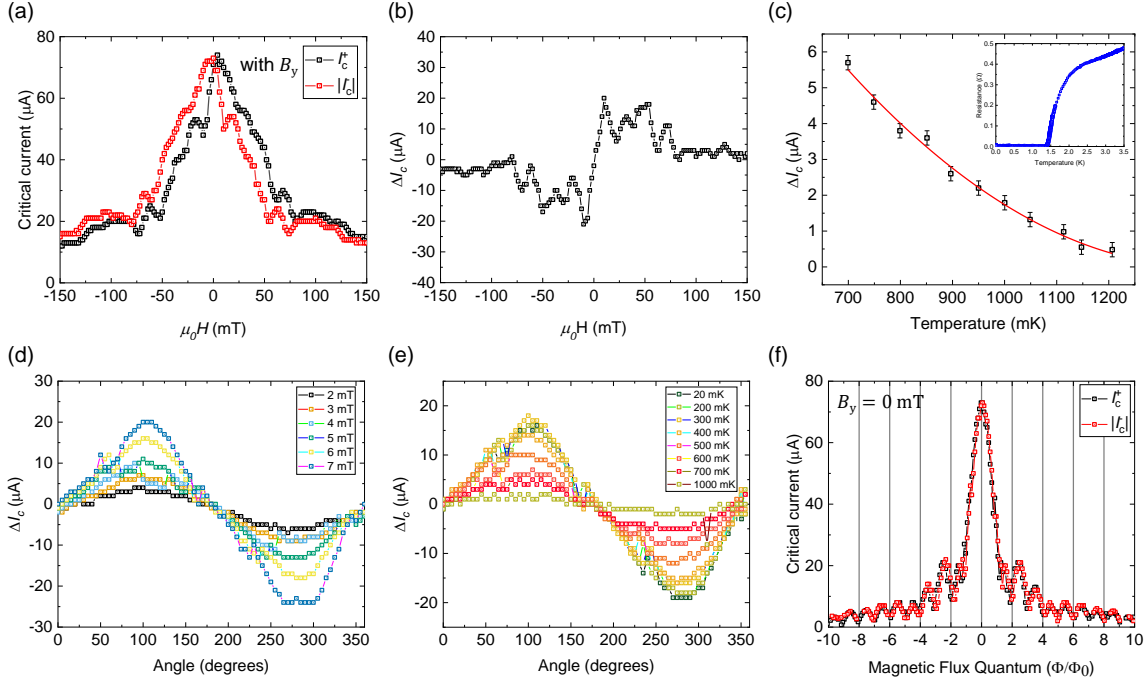

**Fig. 2 Analysis of $\Delta I_c$ with magnetic field magnitude, angle and temperature for junction L1.** (a) $I_c^+$ (in black) and $|I_c^-|$ (in red) measured as a function of the magnetic field $B_y$ swept from 150 mT to −150 mT at 20 mK temperature. (b) JDE ($\Delta I_c$) measured by sweeping $B_y$ from 150 mT to −150 mT, shows that it is maximum around 10 mT. (c) This figure shows the temperature dependence of $\Delta I_c$ measured in device B1 at higher temperatures and $B_y = 24$ mT. The fit represents a quadratic $(T - T_J)^2$ dependence with $T_J \approx 1.4$ K. Inset shows the measurement of resistance vs temperature in the presence of a magnetic field $B_y = 24$ mT, which shows $T_J$ to be around 1.4 K (d) The angular dependence of $\Delta I_c$ at various magnetic fields measured at 20 mK shows that $\Delta I_c$ is maximized when the magnetic field is perpendicular to the direction of current and zero when the magnetic field is parallel to the direction of current indicating a helical spin-momentum locking in the system. (e) The angular dependence of $\Delta I_c$ with the magnetic field $B_y = 8$ mT measured at various temperatures. (f) $I_c^+$ and $I_c^-$ measured as a function of the magnetic field $B_z$ measured in a criss-crossed configuration shows negligible $\Delta I_c$.

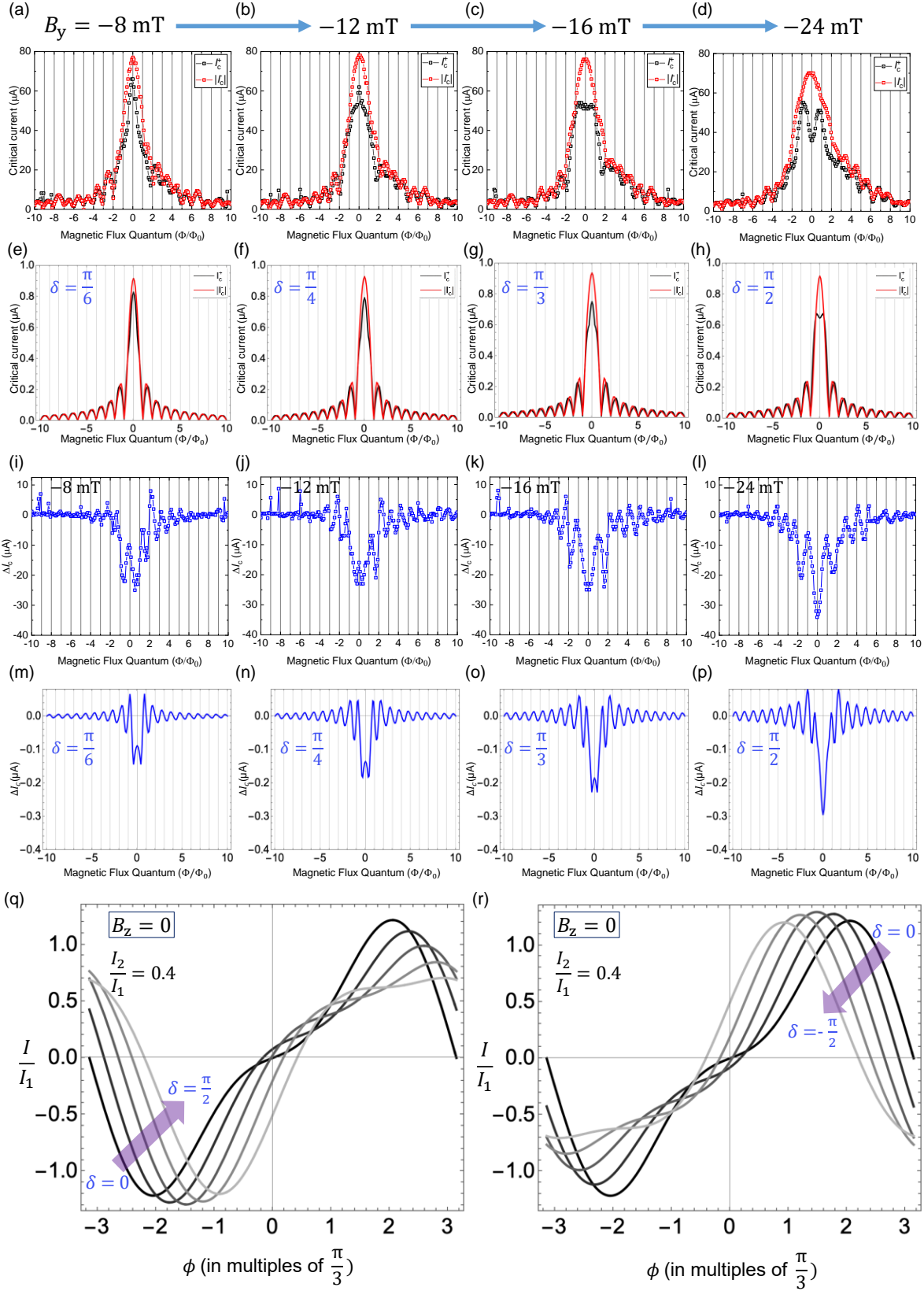

**Fig. 3 Evolution of the Fraunhofer pattern in the presence of $\Delta I_c$ for junction L1.** (a)-(d) shows the experimentally measured Fraunhofer patterns for $I_c^+$ and $I_c^-$ in the presence of a negative $B_y$ of different magnitudes up to $-24$ mT. A 0-π junction-like dip is observed in $I_c^+$ upon increasing the magnitude of $B_y$. (e)-(h) shows the simulated Fraunhofer patterns using a self-consistent treatment (as described in Supplementary Note 3) for $\frac{I_2}{I_1} = 0.4$. A behavior similar to that in experiment with increasing $\delta$ in the CPR is observed. (i)-(l) shows the increasing $\Delta I_c$ with the increasing magnitude of $B_y$ reaching maximum value around $-24$ mT. (m)-(p) Simulated $\Delta I_c$ using the CPR in equation (3) for similar magnetic fields as in experiment. The experimentally observed features including the dip are captured well by the simulation. (q), (r) shows the CPRs corresponding to negative and positive $B_y$ used in the simulations for $\frac{I_2}{I_1} = 0.4$. The non-reciprocal response of $I_c^+$ and $I_c^-$ under $|B_y|$ is evident from these simulations.

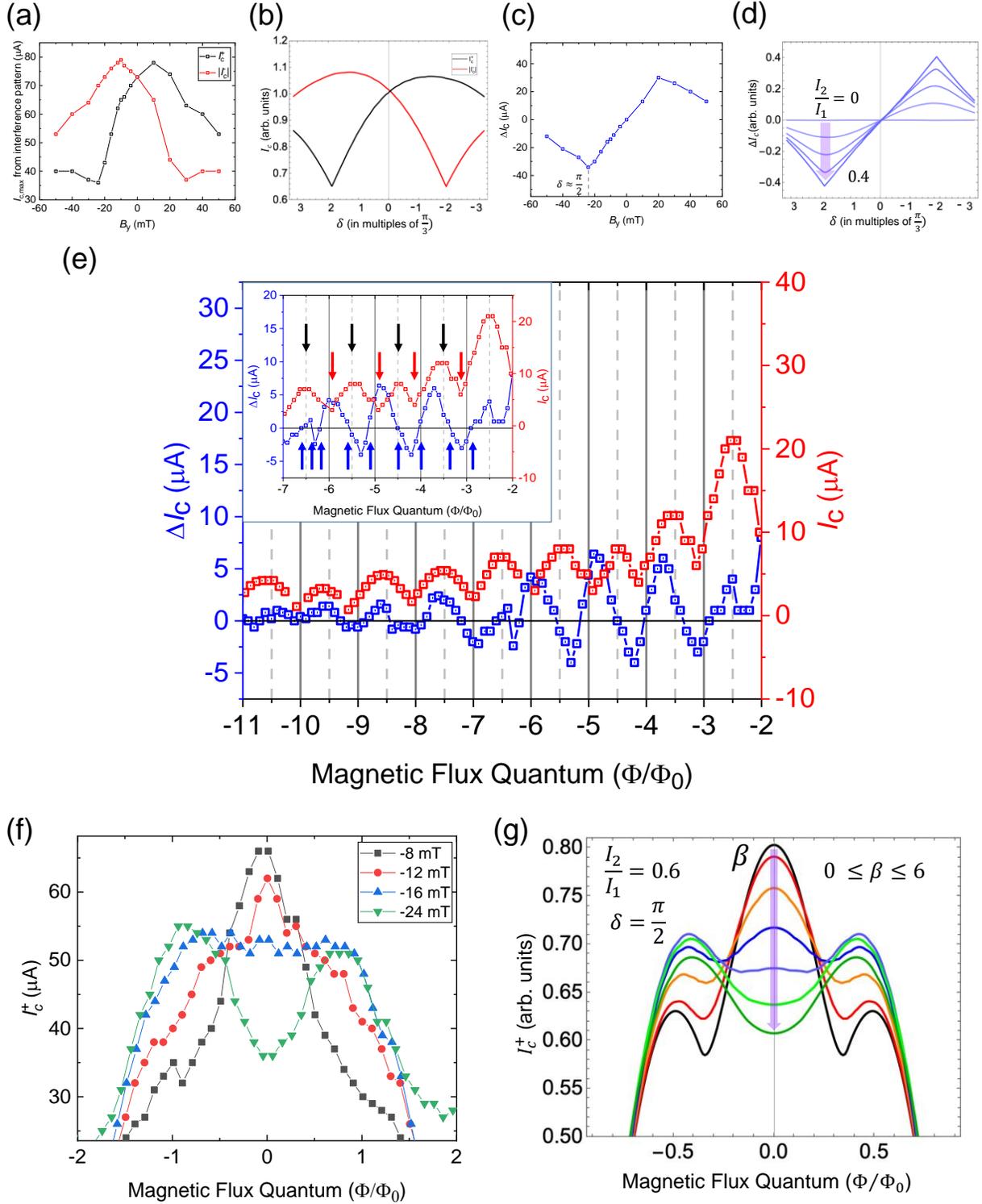

**Fig. 4 Evolution of $\Delta I_c$ with $\delta$ and $\varphi$ in junction L1.** (a) The evolution of $I_c^+$ and $I_c^-$ of the central peak in the Fraunhofer pattern at positive and negative $B_y$ after correcting for finite thickness shifts.

(b) Simulated evolution of $I_c^+$ and $I_c^-$ using the CPR in equation (3). It is found to fairly replicate the experimentally observed features. The absence of nodes in (a) can be replicated by the presence of additional magnetic flux as shown in Supplementary Note 13. (c) $\Delta I_c$ from the Fraunhofer patterns calculated after correcting for finite thickness effects in the junction. (d) $\Delta I_c$ calculated from the simulated Fraunhofer patterns for a wide junction, where $\left|\frac{I_2}{I_1}\right|$ increases in increments of 0.1. The maxima (minima) corresponds to $\delta = -\frac{\pi}{2}\left(\frac{\pi}{2}\right)$. $\Delta I_c$ evolves from a sinusoidal dependence at low values of $\left|\frac{I_2}{I_1}\right|$ to a nearly triangular behavior at higher values of $\left|\frac{I_2}{I_1}\right|$. (e) The evolution of $\Delta I_c$(in blue) and $I_c$(in red) with $\Phi$ with $B_y = 20$ mT. Inset shows a close up of oscillations in $\Delta I_c$ has nodes appearing roughly at half magnetic flux quantum $\left(\frac{\Phi_0}{2}\right)$ frequency(denoted by blue arrows) and has almost double the frequency compared to the nodes in the critical current (denoted by red arrows) that happens roughly at every magnetic flux quantum ($\Phi_0$). It is clearly noticeable that there are twice as many blue arrows compared to red arrows. Black arrows indicate the position of the antinodes in $I_c$. The position of the nodes are slightly altered from half magnetic flux quantum due to the presence of lifted nodes and varying flux focusing factor with increasing magnetic flux. (f) The experimental evolution of $I_c^+$ with $\Phi$ for different values of $\delta$ shows the appearance of a dip with increasing $\delta(-B_y)$. (g) Similar appearance of a dip-like feature in $I_c^+$ is captured in the simulations by tuning the parameter $\beta$. The difference in the values of magnetic flux between the experiment and the simulations can occur due to the presence of a variable flux-focusing factor[23].

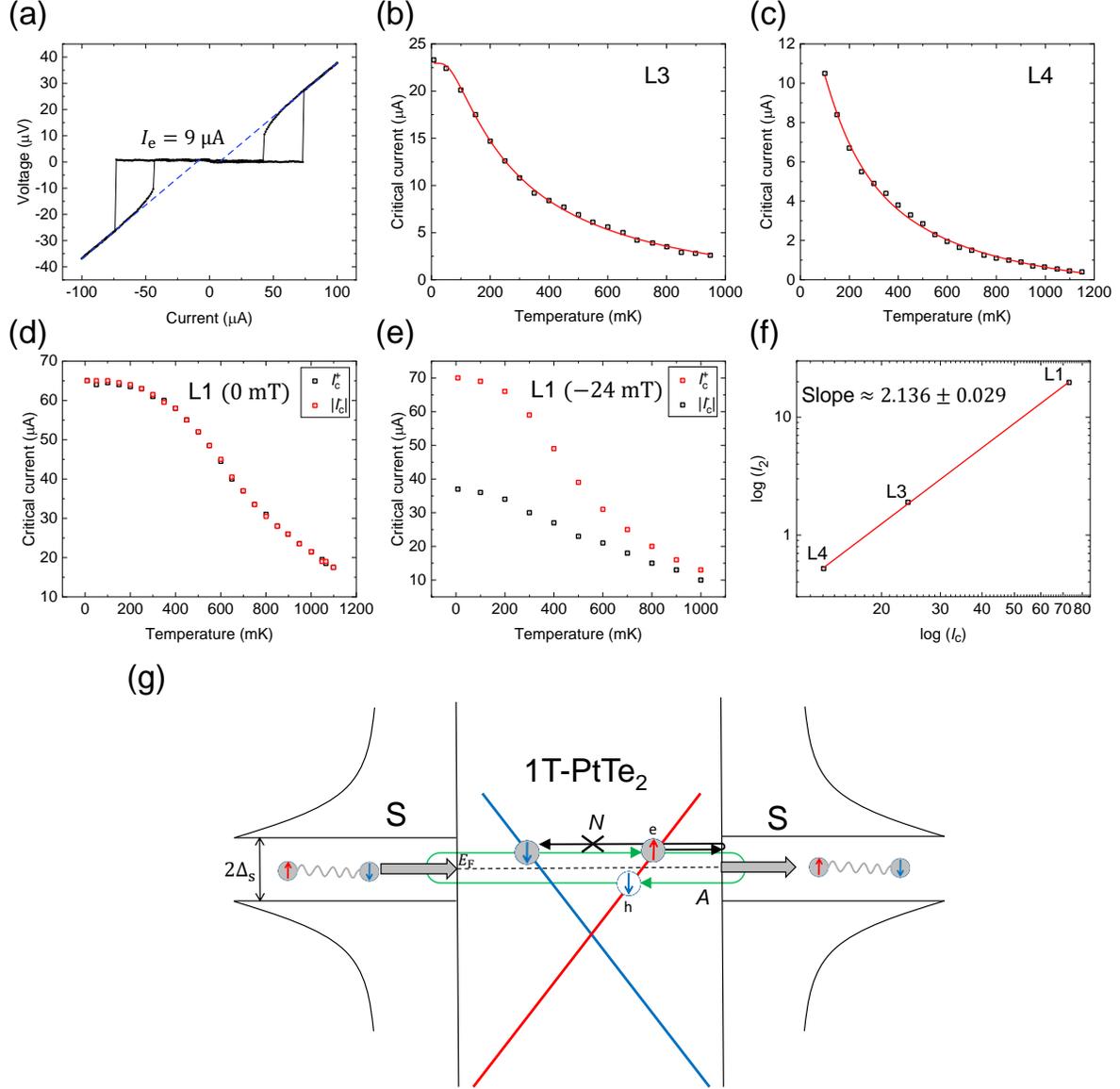

**Fig. 5 Transparency of PtTe$_2$ Josephson junctions.** (a) The $I-V$ curve for L1 junction measured at 20 mK shows the presence of excess currents around 9 μA indicating coherent transport across the junction and a transparency of around 0.45 derived from the OTBK model. (b), (c) $I_c(T)$ for junctions L3 and L4 are fit with an equation corresponding to the long junction limit yielding a transparency of around 0.436 and 0.428 respectively. (d) $I_c(T)$ for junction L1 with $I_c$ starting to saturate below 500 mK. (e) $I_c(T)$ for $I_c^+$ and $I_c^-$ at $-24$ mT for L1 shows that $I_c^-$ has a larger energy gap at low temperatures ($\Delta_0$) in comparison with $I_c^+$ for which $\Delta_0$ is strongly suppressed by the magnetic field. (f) The log-log plot of the evolution of I$_2$ extracted from $\Delta I_c$ with $I_c$ for junctions L1, L3 and L4 shows that the extracted second-harmonic supercurrents scale

quadratically (slope ~ 2) with the critical current as expected[4] ($I_2 \propto I_c^2$, for $I_2 \ll I_1$), further validating the CPR. (g) Schematic of Dirac-cone like helical spin-momentum locking in PtTe$_2$ that potentially suppresses the normal reflections in the junction due to the reduced availability of spin states, The red and blue colors depict opposite spin polarizations in the band. Green arrow represents the Andreev reflection process (*A*) and black arrow represents the normal reflection process (*N*), which is potentially suppressed due to the reduced availability of a state with the same spin. This phenomenon can increase the transparency of the junction, while simultaneously enhancing the phase-coherent Andreev processes, leading to higher harmonics in the supercurrent.

## Acknowledgments

P.K.S. would like to thank Niels Schröter for useful discussion. S.S.P.P. acknowledges support from the European Research Council Advanced Grant SUPERMINT, (project number 101054860). M.T.A. acknowledges support from the NSF through the University of Illinois at Urbana-Champaign Materials Research Science and Engineering Center DMR-1720633.

## Author Contributions

P.K.S. and S.S.P.P. conceived the project. P.K.S. performed the exfoliation and fabrication of lateral junctions. J.-K.K. fabricated the vertical Josephson junctions. Y.W. fabricated the Hall bar devices. P.K.S. performed all the electrical measurements with assistance from A.D. P.K.S. and M.T.A. performed all the data analysis with help from A.K.P. P.K.S. and M.T.A. performed the CPR analysis. M.T.A. performed all the self-consistent CPR simulations with assistance from G.J.C. and M.G. P.K.S., M.T.A., M.G. and S.S.S.P. discussed the data and wrote the manuscript with contributions from all authors.

# Supplementary Information

# Long-range Phase Coherence and Tunable Second Order $\varphi_0$-Josephson Effect in a Dirac Semimetal 1T-PtTe$_2$

Pranava K. Sivakumar[1], Mostafa T. Ahari[2], Jae-Keun Kim[1], Yufeng Wu[1], Anvesh Dixit[1], George J. de Coster[3], Avanindra K. Pandeya[1], Matthew J. Gilbert[2,4] and Stuart S. P. Parkin[1].

5. Max Planck Institute of Microstructure Physics, 06120 Halle (Saale), Germany
6. Materials Research Laboratory, The Grainger College of Engineering, University of Illinois, Urbana-Champaign, Illinois 61801, USA
7. DEVCOM Army Research Laboratory, 2800 Powder Mill Rd, Adelphi, Maryland 20783, USA
8. Department of Electrical Engineering, University of Illinois, Urbana-Champaign, Illinois 61801, USA

# Contents for Supplementary Notes

The supplementary information file contains the following material that supports the data analysis and main conclusions of the manuscript.

1. Thickness of PtTe$_2$ flake

2. Electrical transport properties of PtTe$_2$ in the normal state

3. Self-consistent treatment of the wide Josephson junction

4. Identification and Minimization of JDE induced by self-field effects in wide PtTe$_2$ Josephson junctions

5. Josephson diode effect in a vertical Josephson junction of PtTe$_2$

6. $\Delta I_c$ due to the geometric shape inversion asymmetry in PtTe$_2$ junctions

7. Effect of finite thickness of PtTe$_2$ flake on the interference pattern

8. Flux focusing and Estimation of effective junction area for diffraction pattern calculations

9. Evolution of Fraunhofer patterns in positive and negative magnetic fields for L1

10. $\Delta I_c$ from device L3

11. $\Delta I_c$ from device L4

12. List of all measured junctions with different seperations

13. Effect of magnetic flux on $\Delta I_c$

14. Accidental SQUID in junction L2

15. Josephson diode efficiency in PtTe$_2$ junctions

16. Discussion on $g$-factor estimation from JDE

Supplementary References

## 1. Thickness of PtTe$_2$ flake

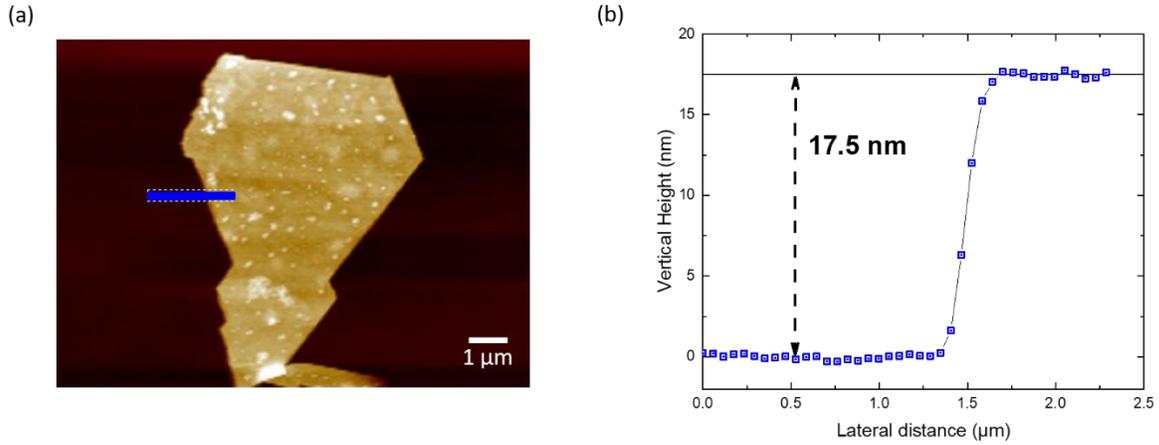

**Fig. S1 AFM image and thickness of PtTe$_2$ flake used.** (a) AFM image of the as exfoliated PtTe$_2$ flake on Si/SiO$_2$ substrate. The scale bar corresponds to 1 μm. (b) shows the height of the PtTe$_2$ flake measured from AFM measurement at the position marked with a blue line in (a).

## 2. Electrical transport properties of PtTe$_2$ in the normal state

A separate Hall bar device with Ti (2 nm) / Au (40 nm) contacts was fabricated on a 30 nm thick PtTe$_2$ flake using a similar method as described above to measure its electrical properties in the normal state as shown in Fig. S2. Electrical measurements are performed using a current source (Keithley 6221) with a constant current bias of 10 μA and the voltage measured using a nanovoltmeter (Keithley 2182A). It is found that PtTe$_2$ remains metallic down to 20 mK temperature and does not turn superconducting. It can also be seen from the positive magnetoresistance and linear Hall effect that PtTe$_2$ is non-magnetic down to the base temperature and hence time-reversal symmetric. The mobility of the flake as estimated from the magnetoresistance is $\mu_{MR} = 835 \text{ cm}^2\text{V}^{-1}\text{s}^{-1}$ and the carrier concentration calculated from the Hall resistance is $n = 1.424 \times 10^{28} \text{ m}^{-3}$.

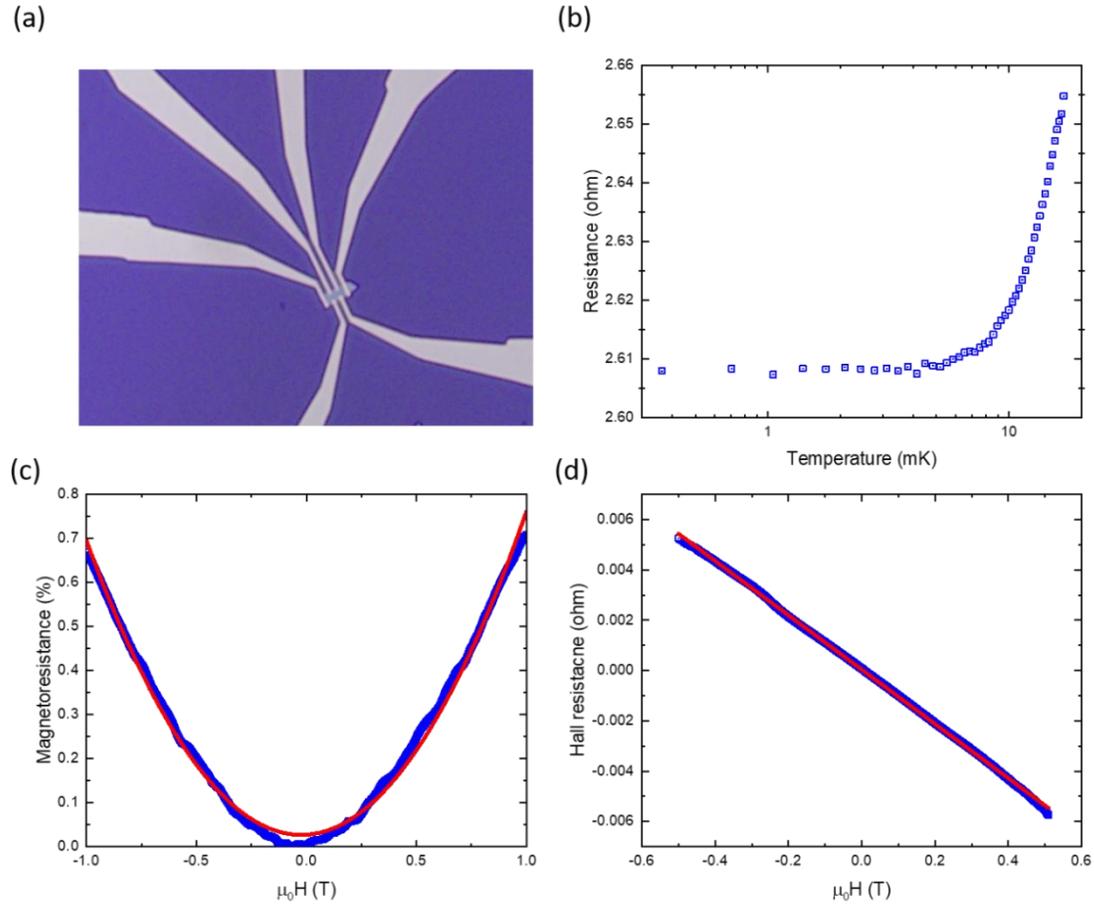

**Fig. S2 Electrical transport properties of an exfoliated PtTe₂ flake.** (a) Optical image of a Hall bar device with gold electrodes fabricated on a $30\ nm$ thick PtTe$_2$ flake exfoliated on Si/SiO$_2$ substrate. (b) Resistance of the PtTe$_2$ flake measured as a function of temperature shows that it remains metallic down to $20\ mK$. (c) Magnetoresistance of the PtTe$_2$ flake is positive and shows quadratic behavior indicating the absence of magnetism in the system. (d) The Hall resistance of the PtTe$_2$ flake is linear with a negative slope indicating the transport is hole-like with a carrier density of $n = 1.424 \times 10^{28}$ m$^{-3}$.

## 3. Self-consistent treatment of the wide Josephson junction

Consider the Josephson junction under a uniform external magnetic field. Since the junction thickness in the z direction $t \approx 17$ nm is much less than the London penetration depth ($\lambda_L$), as result, we expect a large self-inductance influencing the phase $\varphi$, and $B_y$ penetrates the junction without distortion. In the presence of out-of-plane filed $B_z$, the phase $\varphi$ obeys the differential equation

$$\mu_0 J_c \lambda_J^2 \frac{\partial \varphi}{\partial y} = B_z, \tag{1}$$

where $\mu_0$ is the permeability of free space, $J_c \equiv J_1$ is the critical density, $\lambda_J$ is the Josephson penetration length. Since $\lambda_J$ depends on the details of the junction [73-75] as we discuss below, we use different current bias configurations to directly obtain $\lambda_J$. When the current density is sufficiently strong, the supercurrent flow in the junction tends to screen the magnetic field from the interior of the junction[76], and the magnetic field satisfies the Maxwell equation

$$\frac{\partial B_z}{\partial y} = \mu_0 J_x(y) \tag{2}$$

Now, combining Eqs. (1) and (2), we write

$$\frac{\partial^2 \varphi}{\partial y^2} = \frac{1}{\lambda_J^2}\left(\sin \varphi + \frac{J_2}{J_1}\sin 2\varphi + \delta\right) \tag{3}$$

We note that when $J_2 = 0$, Eq. (3) reduces to the usual wide junction limit with a sinusoidal CPR, the so-called (static) sine-Gordon equation[77]. However when $J_2 \neq 0$, Eq. (3) is known as the (static) double sine-Gordon equation. Note that in the main text, when we discuss the CPR we use $I_2 \equiv J_2 tW$, and $I_1 \equiv J_1 tW$. The DC limit of our junction is well captured by solutions of this equation, which corroborates the validity of our CPR. Converting to dimensionless equations, we adopt the boundary conditions[78], for the phase gradient as

$$W \frac{\partial \varphi}{\partial y}\bigg|_{y=-\frac{W}{2}} \approx \alpha 2\pi \frac{\Phi}{\Phi_0} + a_1 \left(\frac{W}{\lambda_J}\right)^2 \frac{I}{I_c}$$

$$W\frac{\partial\varphi}{\partial y}\bigg|_{y=\frac{W}{2}} \approx \alpha 2\pi\frac{\Phi}{\Phi_0} + a_2\left(\frac{W}{\lambda_J}\right)^2\frac{I}{I_c}$$

where $I_c = J_c t W$, $\alpha = \Phi_0/(2\pi\mu_0 d_{eff} J_c \lambda_J^2)$ is a dimensionless constant parameter which we determine from our simulation to be $\alpha \approx 1$, with $d_{eff} = d + 2\lambda_L$, and $I = t\int_{-W/2}^{W/2} J_x(y)dy$ is the total supercurrent flowing in the junction. Here $\Phi$ is the external magnetic field's flux threading the junction. The same-side and opposite-side (criss-crossed) biasing can now be easily adopted by $(a_1, a_2) = (0,1)$ and $(-1/2, 1/2)$, respectively. It is clear that total current affects the solutions of the phase, and, in turn, the phase determines the current density $J_x(y)$. As a result, these equations are solved self-consistently. In order to obtain the Fraunhofer patterns, we find the maximum and minimum total supercurrents, $I_c^+ = \max[I]$ and $I_c^- = \min[I]$, that satisfy the boundary conditions. . At zero in-plane field $B_y = 0$, Fig. S3c and S3f present our simulated Fraunhofer patterns with $\lambda_J = W/2$ for same-side and opposite side biasing, respectively. We point out that, depending on the value of $\frac{J_2}{J_1}$, our simulations with the penetration length within the range $1 < W/\lambda_J < 4$ can match well with the experimental results.

For the opposite-side bias boundary conditions, one can check that a self-consistent (numerical) solution of Eq. (3) produces the main features of Fraunhofer patterns, such as lifted nodes and a local minimum (dip) in $I_c^-$. However, for simplicity we consider a simple solution $\varphi = 2\pi\frac{\Phi}{\Phi_0}\frac{y}{W} + \frac{1}{2}\left(\frac{W}{\lambda_J}\right)^2\frac{I}{I_c}\left|\frac{y-y_0}{W}\right|$ that satisfies the boundary conditions and captures the central local minimum feature in the Fraunhofer patterns. While our simulations in the main text are for $y_0 = 0$, a nonzero $y_0$ can lead to an asymmetry in Fraunhofer patterns as $I_c^\pm(B) \neq I_c^\pm(-B)$.

# 4. Identification and Minimization of JDE induced by self-field effects in wide PtTe₂ Josephson junctions

In a Josephson junction, when the junction width ($W$) is smaller than the Josephson penetration depth $(\lambda_J)$, which is a characteristic length scale in the junction over which magnetic flux variation can take place, the magnetic field due to the current flow through the junction can be neglected. Such a junction is said to be in the 'short junction limit'. In contrast when the width of the junction is larger in comparison to the Josephson penetration depth $(W > \lambda_J)$, the effect of current-induced magnetic field becomes significant. In such limit, the geometry of the current source configuration can play a significant role in the current distribution across the junction. In a lateral junction as here, when the current source is connected to leads on the same side of the device and the voltage probes are connected to the leads on the other side (as in Fig. S3a), this can lead to a non-uniform current distribution in the junction. This consequently creates local inhomogeneous magnetic fields in the junction that can break the time reversal symmetry of the junction. This is known in the literature as a 'self-field effect (SFE)'[79-81] and is dictated purely by the geometry of the junction and the magnitude of the critical current. While measuring the current-phase relationship of the junction by applying a magnetic field perpendicular to the plane of the sample ($B_z$), this SFE can lead to the creation of Fraunhofer oscillations that are skewed to one side and which alternates with the current direction. Such a skewed Fraunhofer pattern has been depicted in Fig. S3c. It is clear that there is a difference in the critical currents that is induced by skewing of this interference pattern, which leads to a $\Delta I_c$ as shown in the inset to Fig. S3c. This $\Delta I_c$ induced by SFE looks very similar to those induced by finite momentum Cooper pairing in the system[82] but arises purely due to the extrinsic SFE and has little to do with the intrinsic properties of the junction material. Hence, it is important to make sure that the skewedness and JDE arising due to such geometric effects are completely nullified while measuring $\Delta I_c$ and deriving conclusions about the material properties based on it. We note that even though the SFE evidently breaks time-reversal symmetry upon application of a current, there is no spontaneous time-reversal symmetry breaking in the absence of a current and the inhomogeneous magnetic fields induced by the currents in positive and negative directions are the same in magnitude. Hence, there is no $\Delta I_c$ at zero magnetic field in the presence of SFE-inducing currents.

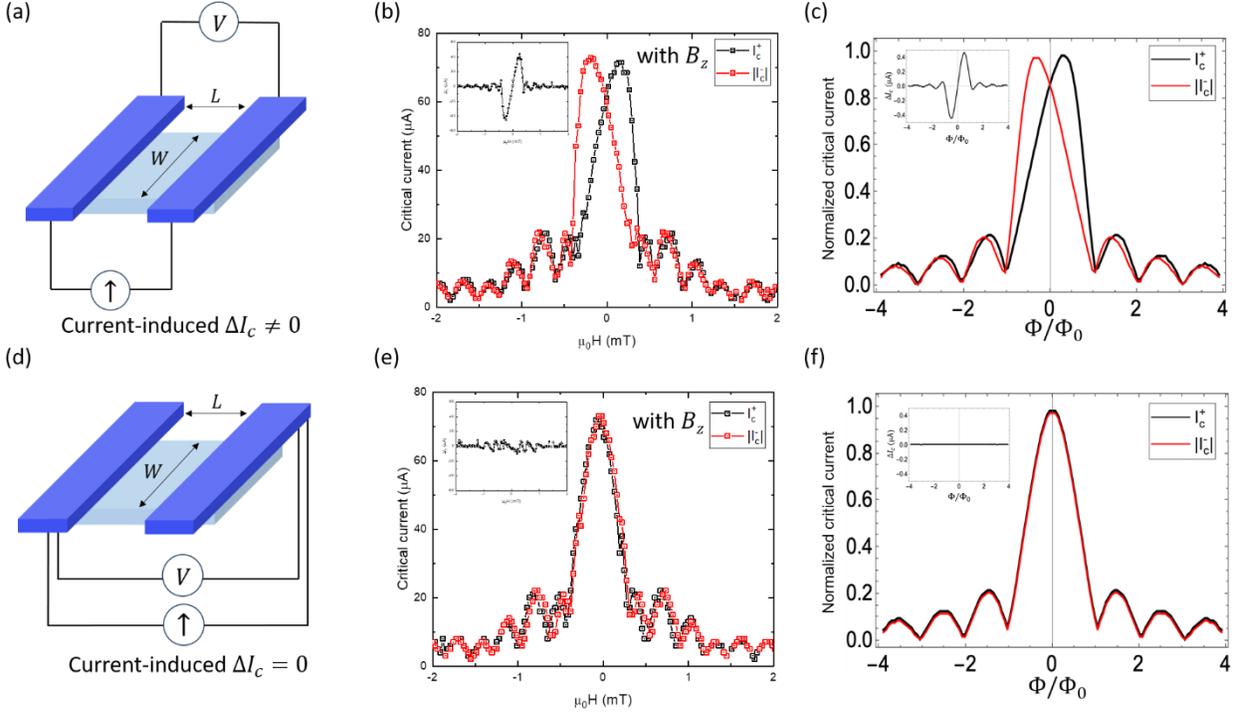

**Fig. S3 Self-field effects in wide PtTe$_2$ junctions.** (a) Measurement schematic for a wide lateral Josephson junction in which self-field effects were observed in the Fraunhofer pattern. (b) Skewed Fraunhofer pattern with large $\Delta I_c$ (inset) measured with the current leads on the same side. (c) Simulated Fraunhofer patterns for same-side biasing with $W \sim \lambda_J$. The self-field effects gives rise to skewed Fraunhofer patterns as in experiments. The calculated $\Delta I_c$ are shown in the inset. (d) Measurement schematic in which the current bias is sourced in the "criss-crossed" configuration, wherein the current leads are located on opposite sides of the superconducting electrodes. (e) The Fraunhofer pattern for $I_c^+$ and $I_c^-$ measured in the criss-crossed configuration is symmetric with respect to the magnetic field with negligible $\Delta I_c$ (inset) indicating the near uniform flow of supercurrents. (f) Simulated Fraunhofer patterns for criss-crossed biasing with $W = 2\lambda_J$. In this case, self-field effects are symmetric and lead to symmetric Fraunhofer patterns where $\Delta I_c = 0$ (inset). In (c) and (f), we plot the normalized critical currents as a function of normalized magnetic flux $\Phi/\Phi_0$.

In L1, the junction channel is between $5 - 6$ μm wide while the estimated $\lambda_J$ for the junction is $\lambda_J \approx 3$ μm for $B_y = 0$, which makes L1 lie in the wide junction limit $(W > \lambda_J)$. The evolution of the critical current in the presence of $B_z$ is first measured in a geometry where the current source is connected to leads located on the same side of the junction as in Fig. S3a. It displays a skewed critical current behavior (Fig. S3b), which is not symmetric with the magnetic field direction and deviates strongly from the expected behavior for Fraunhofer patterns. $I_c^+$ and $I_c^-$ are found to be skewed along opposite directions leading to a large $\Delta I_c$, as shown in the inset to Fig. S3b. The observed $\Delta I_c$ reaches a maximum value of around 40 μA, when the total maximum critical current is only 73 μA with a very large diode efficiency $\left(\eta = \frac{\Delta I_c}{I_c^+ + |I_c^-|}\right)$ of around 47 %. The simulated Fraunhofer pattern for a wide Josephson junction with SFE is also shown for comparison in Fig. S3c. This result shows the presence of a strong SFE and a non-uniform current distribution when the current bias is applied in the geometrical configuration as in Fig. S3a. The $\Delta I_c$ observed has a large magnitude and strongly mimics that created by spin-momentum locking, including the oscillatory behavior with magnetic field[82]. If not analyzed carefully, this would lead to the possible conclusion that there is a Zeeman-type out-of-plane spin-momentum locking in the system that leads to a large $\Delta I_c$ with a magnetic field along z-axis as in Ising superconductors such as 2H-NbSe$_2$[83]. However, this is not true in our case. To remove the skewedness of critical currents, we use a 'criss-crossed geometry' of the current source. In the criss-crossed configuration (Fig. S3d), for which the current source is connected to the leads on opposite sides of the device, the skewed nature of the Fraunhofer pattern vanishes completely and $I_c^+$ and $I_c^-$ fall on top of each other, as shown in Fig. S3e with a negligible $\Delta I_c$ (in the inset to Fig. S3e). The Fraunhofer pattern displays periodic oscillations of the critical current and is quite symmetric with respect to the direction of magnetic field. This indicates that the distribution of supercurrents, when the current bias is applied in the criss-crossed configuration, is rather uniform and there is negligible extrinsic JDE in the system due to the junction geometry.

## 5. Josephson diode effect in a vertical Josephson junction of PtTe$_2$

A vertical Josephson junction (V1) of a 60 nm thick PtTe$_2$ flake was fabricated using NbSe$_2$ as the superconducting electrode on the top and bottom using a dry transfer technique with a polycarbonate (PC) film coated on a dome-shaped polydimethylsiloxane (PDMS) stamp as described in the literature[84], to look for a Josephson diode effect ($\Delta I_c$) by passing supercurrents along the c-axis (as shown in Fig. S4a). The vertical heterostructure formed is then dropped on pre-sputtered gold electrodes at 200 °C and immersed in chloroform to remove any PC residue. The vertical stack is then annealed in vacuum at 300 °C for an hour to improve the electrical contact to the flakes. No apparent $\Delta I_c$ was observed, when a magnetic field was applied in the plane of the flakes along different directions as shown in Fig. S4b and Fig. S4c even though the Josephson energy and the magnitude of the maximum critical current is quite similar in both the L1 and V1 junctions. This demonstrates that possibly there is no spin-momentum locking in the bulk of the sample and the observation of a $\Delta I_c$ in the presence of a magnetic field is confined to current flow along the two-dimensional *ab*-plane of the sample.

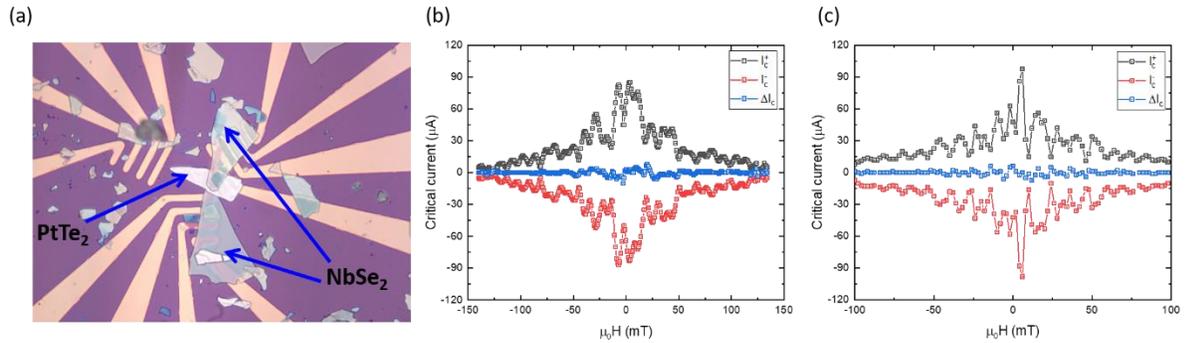

**Fig. S4 Absence of Josephson diode effect in a vertical junction of PtTe₂.** (a) Optical image of a vertical PtTe₂ Josephson junction with superconducting NbSe₂ electrodes in the top and bottom. (b) and (c) Critical currents $I_c^+$ (black) and $I_c^-$ (red) of the vertical junction measured as a function of the in-plane magnetic field at 20 mK with magnetic field applied along two perpendicular directions show that $\Delta I_c$ (blue) is almost zero and has no apparent trend.

# 6.   $\Delta I_c$ due to the geometric shape inversion asymmetry in PtTe₂ junctions

The trapezoidal shape and tapering edges of the PtTe₂ flake in Josephson junctions L1-L4 naturally break the inversion symmetry of the system and might be considered as the origin for the observed $\Delta I_c$ in these junctions. If this is indeed the case, then junctions L1 and L3, which taper in opposite directions should show opposite signs of $\Delta I_c$ when the current and magnetic field are applied along the same direction in both these devices. We show below that this is not the case and we observe $\Delta I_c$ for L1 and L3 is of the same sign when measured under the same conditions, hence ruling out the likelihood of geometric asymmetry contributing significantly to the observed $\Delta I_c$.

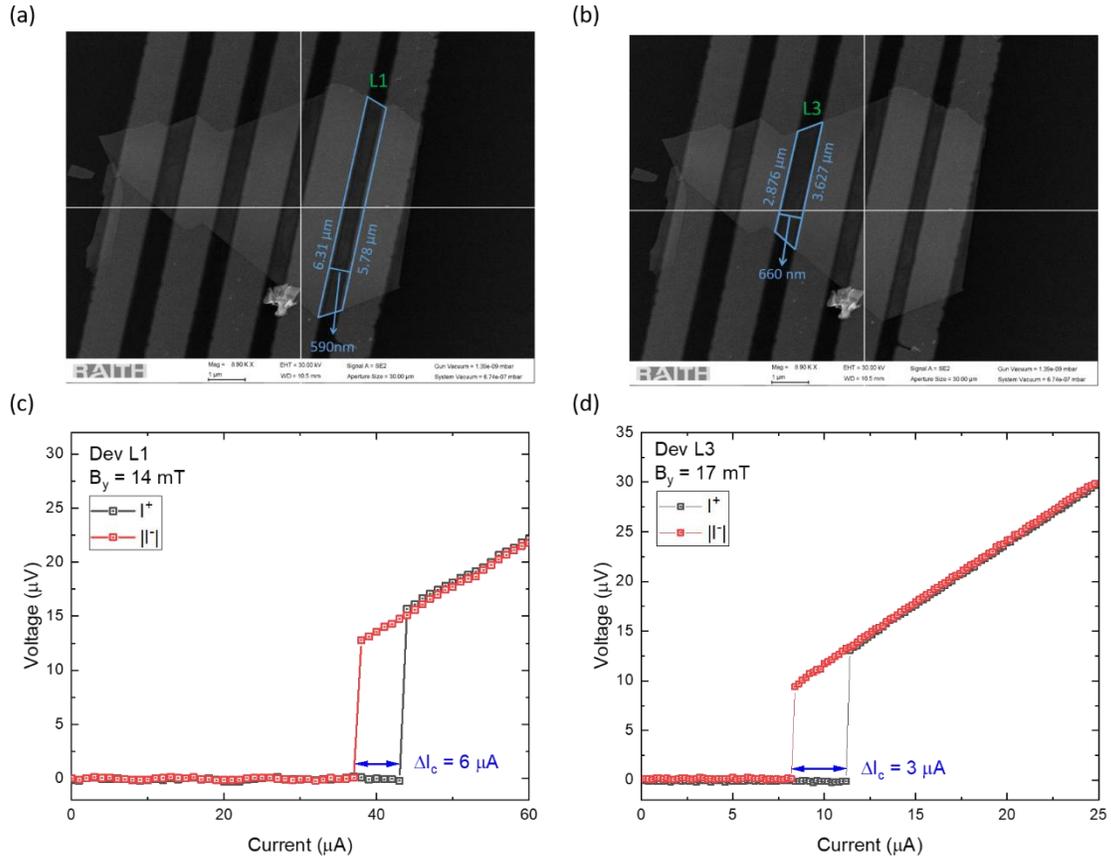

**Fig. S5 Josephson diode effect in junctions of different geometry.** (a) and (b) Scanning electron microscope (SEM) image of lateral PtTe$_2$ Josephson junctions L1 and L3 (described in the main text) with an outline of their trapezoidal shapes and a measurement of their dimensions. (c) and (d) $\Delta I_c$ for L1 and L3 measured for a positive $B_y$ with the same direction of current bias show that both these junctions have the same sign of $\Delta I_c$ under these conditions.

# 7. Effect of finite thickness of PtTe₂ flake on the interference pattern

When the critical current of the junction is measured as a function of the magnetic flux along the z-axis ($B_z$) which induces a phase difference between the superconducting electrodes, we see the expected Fraunhofer interference pattern. When the Fraunhofer pattern is mapped as a function of different in-plane magnetic fields along the y-axis ($B_y$), we observe that there is a uniform shift of the whole Fraunhofer pattern along the $B_z$ with increasing $B_y$, which can be identified by tracking the position of the central maxima. This shift can be effectively modeled by replacing $B_z$ with $B_z + \gamma B_y$, where $\gamma$ is a small fitting parameter introduced when $B_y$ deviates from the y-direction. This shift has also been observed previously as a tilt of the entire Fraunhofer map in similar measurements used to estimate finite-momentum of the Cooper pairs in Josephson junctions[82,85,86]. This tilt is then corrected by subtracting a linear slope that brings the central Fraunhofer maxima back to zero $B_z$.

In our measurements to track the evolution of $I_c^+$ and $I_c^-$ in the Fraunhofer oscillations with $B_y$, we employ a similar procedure to correct for the observed shift of the Fraunhofer pattern by fixing the position of the central maxima at $B_z = 0\ mT$. Below we show the Fraunhofer patterns as measured at different values of $B_y$ (Fig. S6a-d) and after performing the shift correction (Fig. S6e-h) for junction L1. When $B_y$ is increased beyond 30 mT it becomes hard to track the central peak, so the slope of the peak shift with $B_y$ at lower values of $B_y$ can be used to do the shift correction. All Fraunhofer pattern analyses presented in the main text are done after performing this correction.

We note that, despite the correction via a uniform shift, the Fraunhofer patterns are observed to be asymmetric with respect to $\Phi$. We believe this asymmetry is due to the geometric asymmetry in the width of the two leads[86], i.e., $W_1$ and $W_2$ (the trapezoidal shape of the junction).

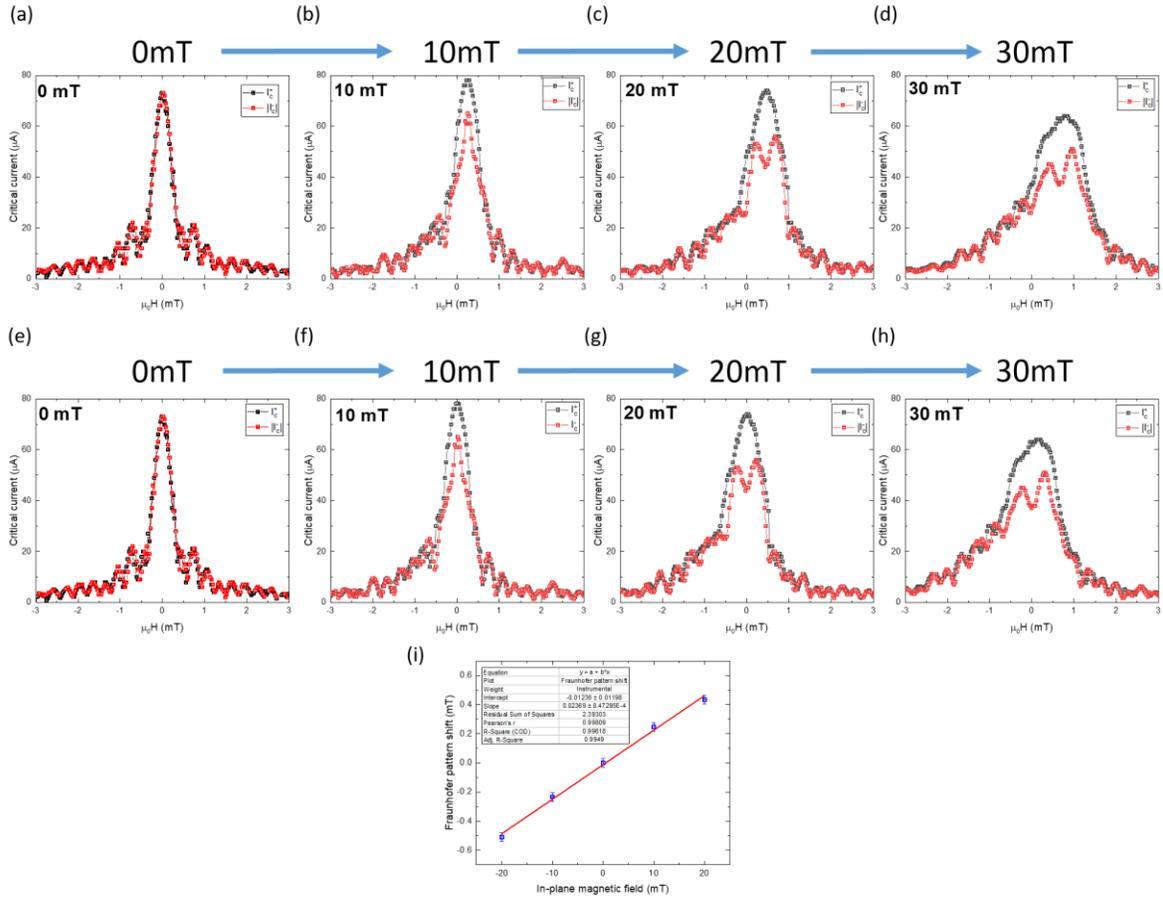

**Fig. S6 Evolution of Fraunhofer pattern for L1 under an in-plane magnetic field and shift correction.** (a)-(d) The Fraunhofer patterns for both $I_c^+$ and $I_c^-$ are found to shift towards the right side along positive $B_z$ values for increasing positive values of $B_y$ and similarly along negative $B_z$ values for negative values of $B_y$. (e)-(h) The Fraunhofer patterns for different positive $B_y$ values after performing the shift correction setting the central maxima of $I_c^+$ to be at $B_z = 0$ mT. (i) Plot showing the linear shift of the Fraunhofer pattern with an in-plane magnetic field $B_y$.

# 8. Flux focusing and Estimation of effective junction area for diffraction pattern calculations

In order to properly estimate the period of oscillations in the observed diffraction pattern, it is important to precisely calculate the effective area of the junction through which the Josephson current flows and the effective magnetic flux through the junction. First, the effective area of the junction including the London penetration depth is calculated and then the effective flux through the junction including flux focusing effects is calculated.

Since the junctions on the PtTe$_2$ flake are shaped like a trapezoid, which is a regular quadrilateral, it makes the calculation of the effective junction area easier. While calculating the area of the junction, it is important to take into account the London penetration depth ($\lambda$) of the niobium electrodes that can increase the effective separation between the two superconducting electrodes. The London penetration depth for thin films of niobium can vary from 37 nm, which is the bulk value up to 200 nm for different thicknesses and temperatures[87-89]. We use a $\lambda$ value of ~ 100 nm as reported in literature[87] for films of thicknesses used in our junctions. That would make the effective junction separation to be $d_{\text{eff}} = (2\lambda + d)$. Then the area of the junction is calculated by using the formula for area of a trapezium $A = \left(\frac{a+b}{2}\right) \cdot d_{\text{eff}}$, where $a$ and $b$ are lengths of the edges of the junction along the electrodes. Using the values estimated from the SEM image in Fig. S5a, this would give an effective junction area of around 3.5665 µm$^2$ for L1. If we use this as the area of the junction, the magnetic flux through the junction can be calculated as $\Phi = B_z \cdot A$ and this magnetic flux normalized to the magnetic flux quantum $\left(\Phi_0 = \frac{h}{2e}\right)$ would be $\left(\frac{\Phi}{\Phi_0}\right)$. The Fraunhofer pattern in these units in presented in Fig. S8a. As it can be observed, the period of oscillations seem to be around $\left(\frac{\Phi_0}{2}\right)$. Ideally, this would only be possible in the case where the second harmonic term in the current-phase relationship (CPR) is the dominant term and the first harmonic component is virtually non-existent in the junction. The expected Fraunhofer patterns for various ratios of the second and first-harmonic components $\left(\frac{I_2}{I_1}\right)$ is presented in Fig.S7. It can be seen that in order to obtain prominent second harmonic oscillations, as we do in our measurements the second-order term ($I_2$) needs to much larger than the first-order term ($I_1$). However, we argue that this is not the case in our junctions as the second-order term stems as a perturbation only because of high transparency in the junction and can't be larger than the first-order term. We also argue that the observed period is due to the effect of flux focusing that is very well known to occur in lateral Josephson junctions.[90,91]

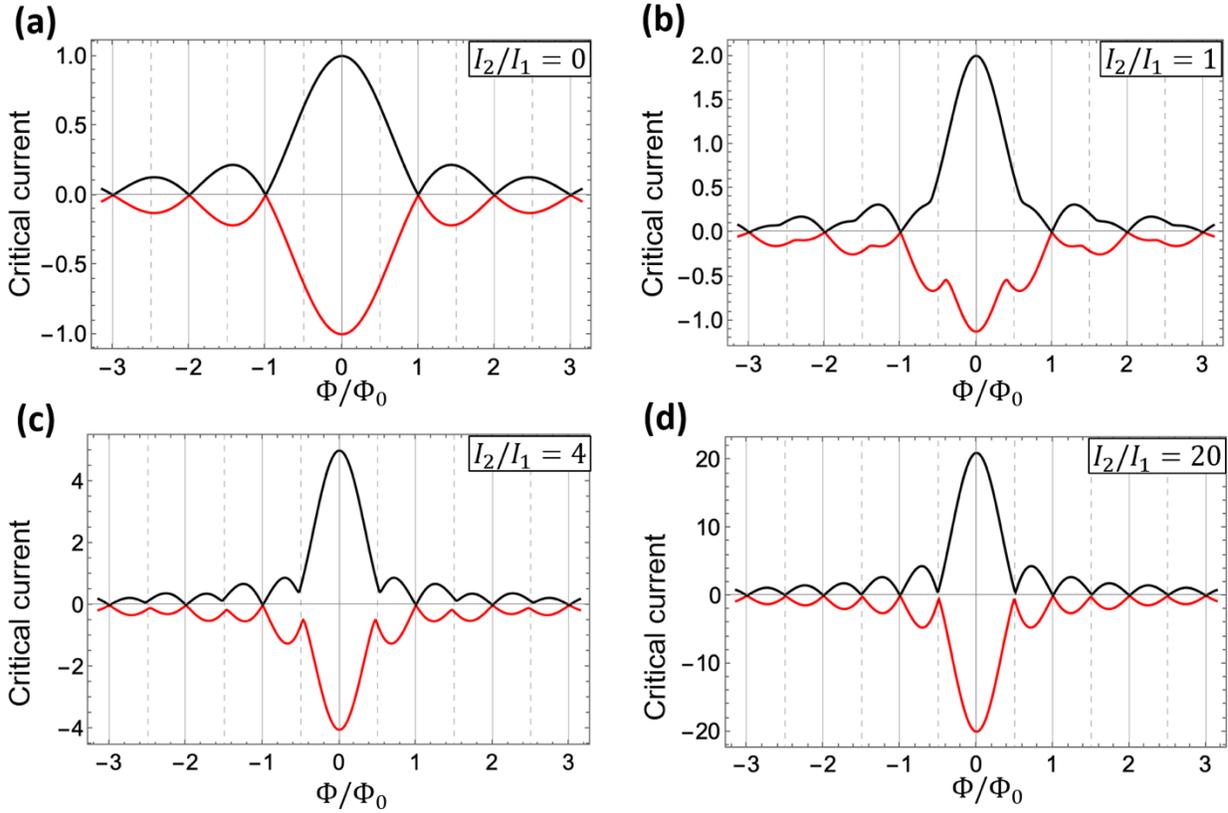

**Fig. S7 The calculated Fraunhofer interference pattern for various values of $I_2/I_1$.** (a) The simulated Fraunhofer patterns expected for various values of $I_2/I_1$ with $\delta = 0$. It can be seen that in order to get prominent $\left(\frac{\Phi_0}{2}\right)$ −periodic oscillations as we observe in our measurements, the ratio of $I_2/I_1$ needs to be very large.

When a magnetic field is applied to a lateral Josephson junction, the Meissner screening currents in the superconducting electrodes deflect a portion of the magnetic flux towards the junction that results in an increased effective magnetic flux than that expected. This is known as the flux focusing effect, which modifies the expected spacing of nodes in Fraunhofer pattern from $\Phi$ to some $\Gamma\Phi$, where the flux scaling factor $\Gamma$ is given by[90]:

$$\Gamma = \frac{n\Phi_0}{B_z^{(n)} LW},$$

where $L = 590$ nm, $W \approx 6$ μm are the junction length and width (for junction L1). $B_z^{(n)}$ is the out of the plane magnetic field at node $n$. A uniform spacing of the nodes are observed as the niobium electrodes are deep in the superconducting state at 20 mK and the Meissner screening effects are constant over the scanned range of $B_z$. For $n = 1$, we have $B_z^{(n)} = 0.3$ mT, which gives $\Gamma \approx 1.9$ that accounts for the observed $\sim\left(\frac{\Phi_0}{2}\right)$ period of the oscillation in the measurements. The Fraunhofer pattern after flux focusing correction is presented in Fig. S8b. The period of oscillations matches well with the expected $\Phi_0$ after accounting for flux focusing effects.

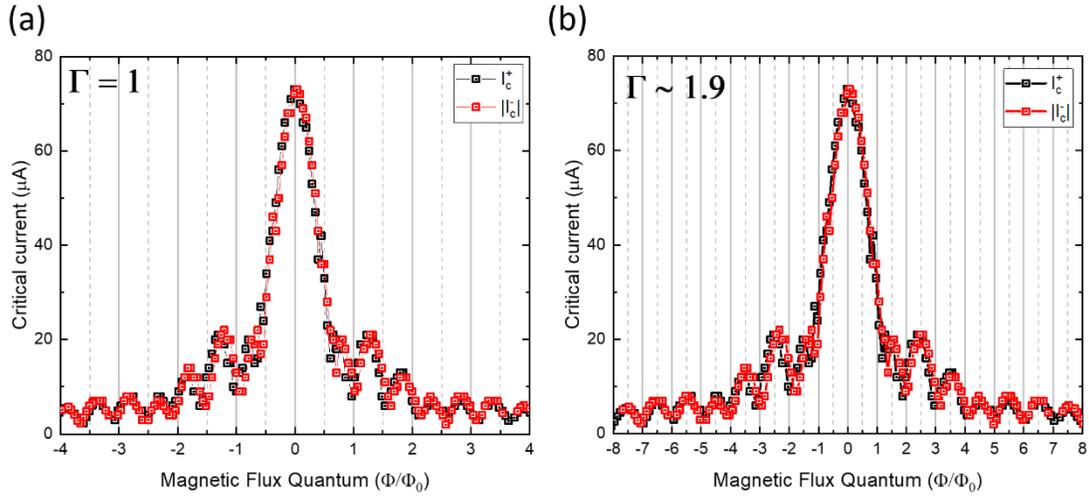

**Fig. S8 The Fraunhofer interference pattern for L1 under before and after correcting for flux focusing.** (a) The as-plotted Fraunhofer patterns for both $I_c^+$ and $I_c^-$ are found to have a $\left(\frac{\Phi_0}{2}\right)$ period before flux focusing correction ($\Gamma = 1$). (b) The Fraunhofer pattern after correcting the applied magnetic flux with the calculated flux scaling factor ($\Gamma \sim 1.9$) matches well with the expected $\Phi_0$ period.

# 9. Evolution of Fraunhofer patterns in positive and negative magnetic fields for L1

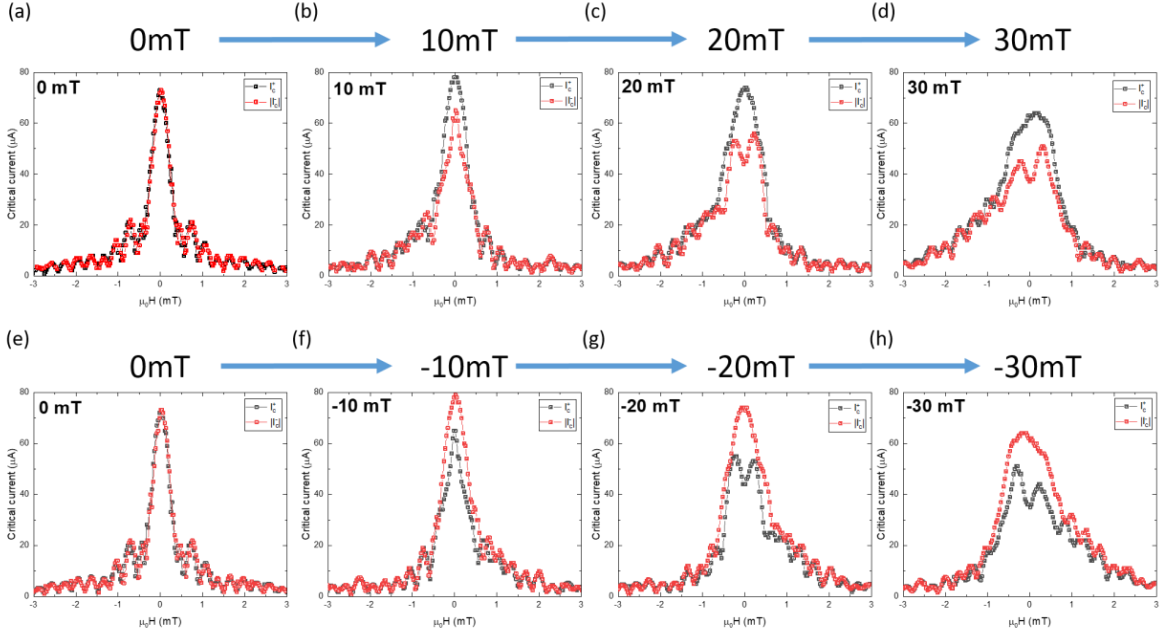

**Fig. S9 The Fraunhofer interference pattern for L1 under positive and negative magnetic fields after shift correction.** (a)-(d) The Fraunhofer patterns for $I_c^+$ and $I_c^-$ measured in the presence of a positive $B_y$ magnetic field as shown in the main text. (e)-(h) The Fraunhofer pattern for $I_c^+$ and $I_c^-$ measured in the presence of a negative $B_y$ magnetic field. The behavior of $I_c^+$ and $I_c^-$ are reversed under opposite $B_y$ but maintain the symmetry $I_c^\pm (B_y, B_z) = I_c^\mp (-B_y, -B_z)$.

# 10. $\Delta I_c$ from device L3

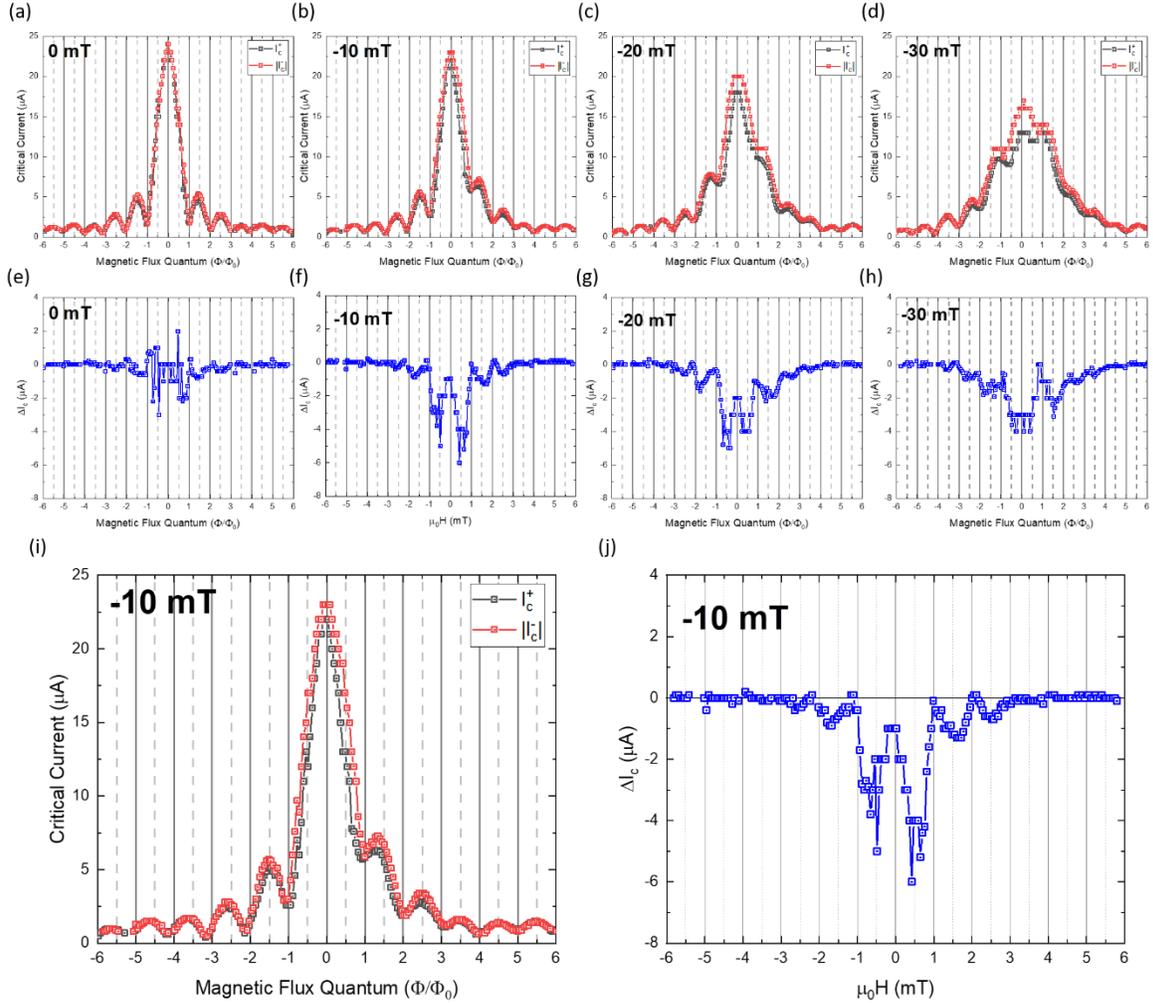

**Fig. S11 Evolution of the Fraunhofer pattern in the presence of $\Delta I_c$ for L3.** (a)-(d) The Fraunhofer patterns for both $I_c^+$ and $I_c^-$ for L3 with increasing $B_y$. (e)-(h) Corresponding $\Delta I_c$ for the Fraunhofer patterns. $\Delta I_c$ for L3 is much smaller compared to L1 as the critical current is also smaller but the $\left(\frac{\Phi_0}{2}\right)$ period of the oscillations can still be distinguished. (i),(j) Larger image of the $-10$ mT data with only a couple of oscillations visible in $\Delta I_c$.

# 11. $\Delta I_c$ from device L4

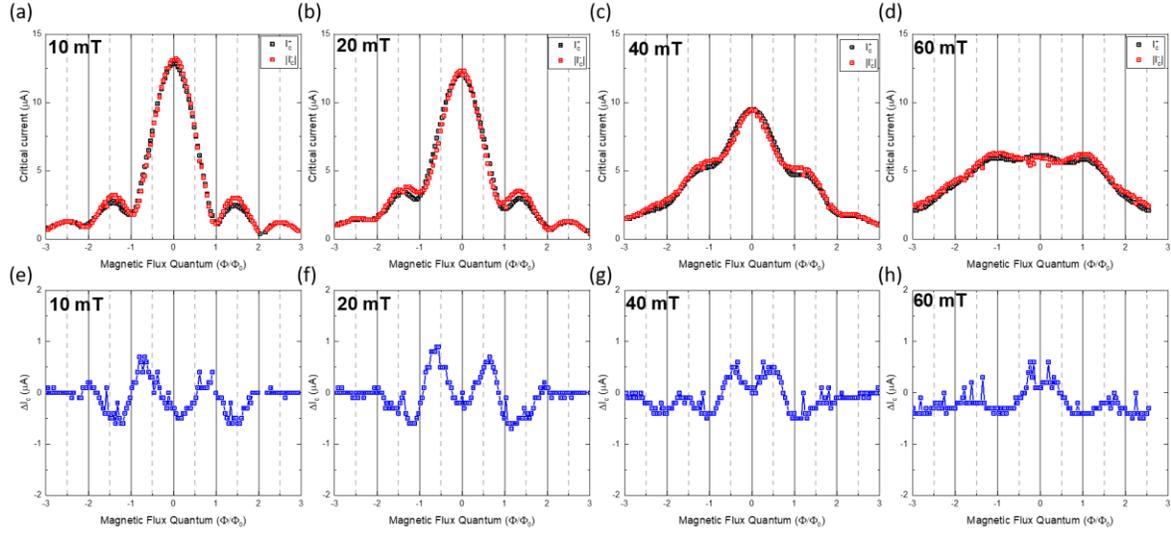

**Fig. S12 Evolution of the Fraunhofer pattern in the presence of $\Delta I_c$ for L4.** (a)-(d) The Fraunhofer patterns for both $I_c^+$ and $I_c^-$ for L4 with increasing $B_y$. (e)-(h) Corresponding $\Delta I_c$ for the Fraunhofer patterns. $\Delta I_c$ in this case is very small and the oscillations are barely discernible.

# 12. List of all measured junctions with different separations

| PtTe$_2$ device | Separation $d$ (nm) | Critical current $I_c$ (µA) | $R_N$ (Ω) | $I_c R_N$ (µV) | Second harmonic $I_2$ (µA) | Transparency $\tau$ |
|---|---|---|---|---|---|---|
| L1 | 390 | 73 | 0.37 | 27.156 | 19.7 | 0.45 (from $I_e$) |
| L3 | 466 | 24 | 1.2 | 28.8 | 1.898 | 0.435 |
| L4 | 597 | 13.4 | 1.72 | 23.04 | 0.52 | 0.427 |

**Table S1 Table of parameters for all measured junctions.** This table contains the critical currents, resistances, transparencies and extracted second harmonic components from the JDE for all junctions on the PtTe$_2$ flake.

## 13. Effect of magnetic flux on $\Delta I_c$

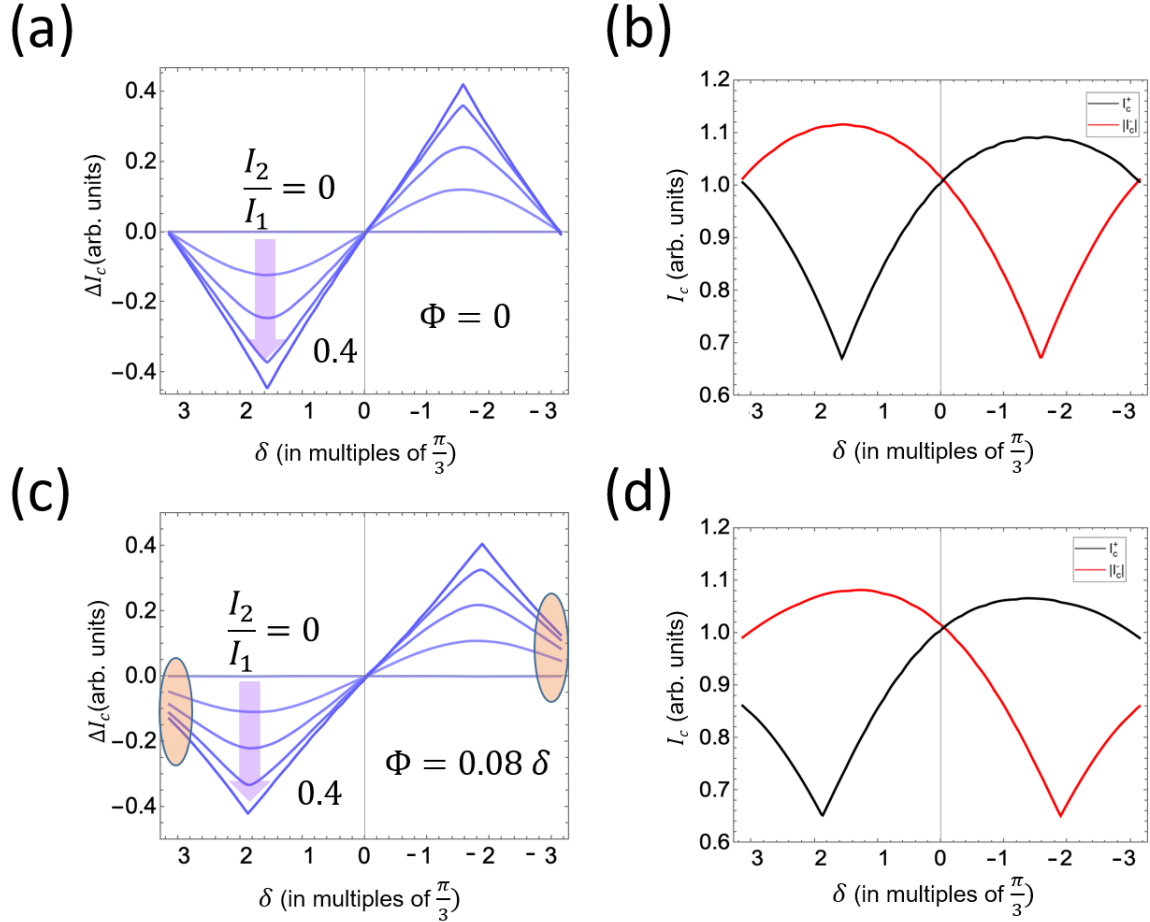

**Fig. S13 Evolution of $\Delta I_c$ in the presence of a small magnetic flux ($\Phi$).** (a), (b) show the simulated behavior of $\Delta I_c$, $I_c^+$ and $I_c^-$ using the CPR in equation (3) of the main text. (c),(d) show the simulated behavior of $\Delta I_c$ in the presence of a magnetic flux ($\Phi$) inside the sample that can modify the behavior and lead to a shift in the position of the nodes in $\Delta I_c$ expected from the CPR. The figures are simulated with an additional flux of $\Phi = 0.08\,\delta$, $\delta \to 0.8\,\delta$ and $\beta = 3$. The circled areas display the lifted nodes. The evolution of $\Delta I_c$ shown in (c) with a shift in the position of the node in $\Delta I_c$ from occurring at $\delta = \pm\pi$ is consistent with the experimental observation in Fig. 4c, indicating the presence of a small flux resulting from the application of $B_y$.

# 14. Accidental SQUID in junction L2

(a) 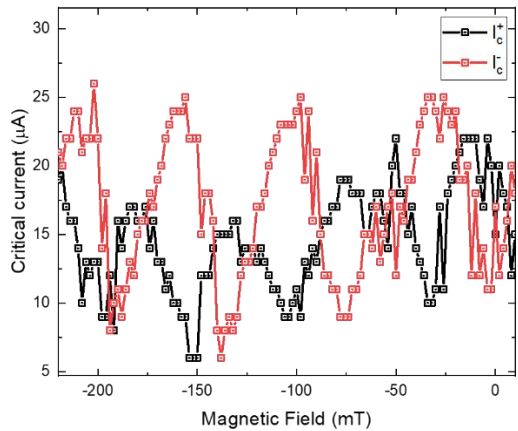

(b) 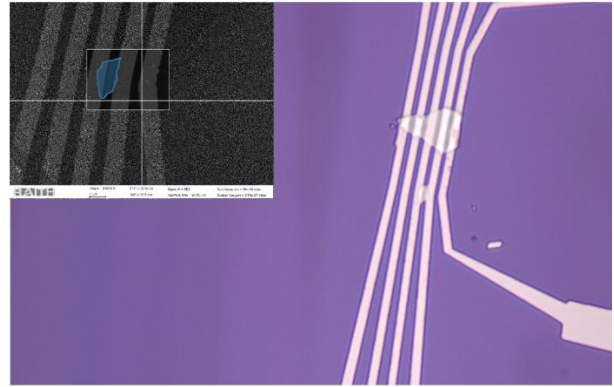

**Fig. S14 Accidental SQUID in junction L2.** Junction L2 is shorted by another flake of PtTe$_2$ by accident forming an asymmetric SQUID loop. (a) The asymmetric SQUID also shows non-reciprocal critical currents with $B_y$ and highly skewed non-sinusoidal oscillations providing further evidence of the presence of higher harmonics in the CPR. (b) Optical image of the shorted junction. Inset shows a close-up SEM image of the flake shorting L2.

## 15. Josephson diode efficiency in PtTe₂ junctions

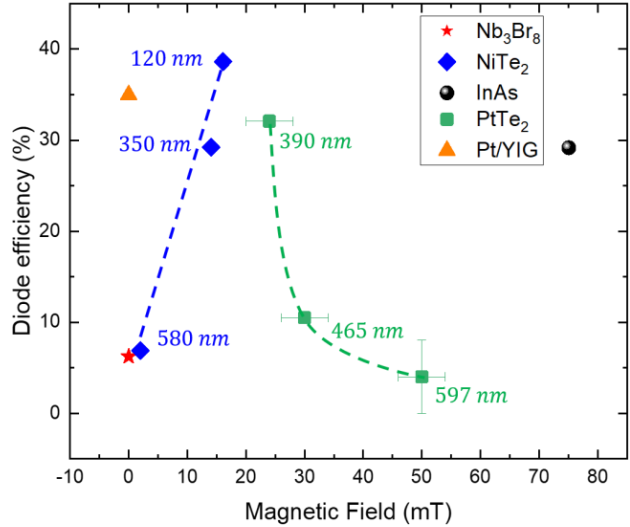

**Fig. S15 Evolution of Josephson diode efficiency in PtTe₂.** The efficiency for various Josephson junction based diodes are presented[82,92-94], including the current work calculated using the formula $\eta = \frac{\Delta I_c}{I_c^+ + |I_c^-|}$ is presented. The efficiency of $\Delta I_c$ in PtTe₂ goes up drastically with decreasing separation and increasing the width of the PtTe₂ junctions.

## 16. Discussion on *g*-factor estimation from JDE

An accurate theoretical estimate of the $g$-factor requires first-principles calculations of the band structure, which would account for both orbital and multiband contributions to the $g$-factor. For the purpose of this discussion, assuming the Fermi velocity of the bands contributing to superconductivity is $v_F \approx 3.3 \times 10^5$ ms$^{-1}$ from literature[95], we can estimate the $g$-factor of the electrons in PtTe₂ from the Cooper pair momentum. The Cooper pair momentum due to the Zeeman effect is given by: $2q = \frac{2g\mu B_y}{\hbar v_F}$ [82,85,86] The diode effect is maximized when $\delta = \frac{\pi}{2}$, that is $2q \cdot d_{eff} = \frac{\pi}{2}$, which in case of device B1 occurs when $B_y = 24 mT$. This gives the Cooper pair momentum to be: $2q \approx 2.345 \times 10^6 \; m^{-1}$ when $B_y = 24 mT$. From this, the value of $\left(\frac{g}{v_F}\right)$ can be estimated to be around $6.3 \times 10^{-4} \; m^{-1}s$. This estimate indicates that the $g$-factor is around 208, which is a rather large number and one of the largest values reported to date. However, it should be noted that this estimate depends on the assumed value of $v_F$ and the accurate determination of $d_{eff} = d + 2\lambda$ for the junction, since $g = \frac{\hbar v_F \pi}{4 d_{eff} \mu B_y}$.

In our junctions, the thickness of the superconducting leads is comparable to the coherence length of niobium, which means that $\lambda$ in our superconducting wire geometry can be significantly larger than that of the $\lambda$ of thin film niobium reported in literature. In our calculation of $\left(\frac{g}{v_F}\right)$, we have assumed $\lambda = 100$ nm from literature. Therefore, $\left(\frac{g}{v_F}\right) \approx 6.3 \times 10^{-4} \; m^{-1}s$ represents rather an upper limit estimate, for the assumed $v_F$. If we were to assume, for instance, $\lambda$ of the superconducting electrodes to be around 500 nm instead, this would result in a $g$-factor of 88. Hence, there is a very large uncertainty in the estimation of $g$-factor from this method, which depends on the accurate estimation of $\lambda$, which in turn depends strongly on various parameters like the exact thickness, geometry and method of deposition of the superconducting electrodes. The estimation of the exact value of $\lambda$ is further complicated by flux focusing effects in the junction.

Though we cannot precisely estimate the $g$-factor in this material from the JDE, we would like to note that there have been recent experiments that have demonstrated the existence of extremely large spin-orbit coupling effects in PtTe$_2$. It has been shown to possess a very large spin Hall angle and spin-orbit torques that are much larger than other topological semimetals like WTe$_2$, Bi$_2$Se$_3$ comparable to metallic Pt[96,97] and has also been used in magnetic switching of ferromagnets[98]. While the presence of large spin-orbit coupling in PtTe$_2$ is clear, how its band structure correlates with its unusually large $g$-factor is something that needs detailed investigation in the future.

# Supplementary References